\documentclass[10pt]{article}
\usepackage[utf8]{inputenc}
\usepackage[T1]{fontenc}
\usepackage{amsmath, amssymb, amsfonts}
\usepackage{geometry}
\usepackage{graphicx}
\usepackage{physics}
\usepackage{booktabs}
\usepackage{subcaption}
\usepackage[
    colorlinks=true,
    linkcolor=blue,
    citecolor=blue,
    urlcolor=cyan
]{hyperref}
\usepackage{amsthm}
\usepackage{tikz}
\usetikzlibrary{decorations.pathreplacing}
\usepackage{algorithm}
\usepackage{caption}
\usepackage{tabularx}
\usepackage{tikz}
\usetikzlibrary{3d,arrows.meta,decorations.markings,shadings}

\usepackage{tikz}
\usetikzlibrary{decorations.pathmorphing, patterns, shapes.arrows, arrows.meta, calc}

\DeclareUnicodeCharacter{21D2}{\Rightarrow}
\DeclareUnicodeCharacter{202F}{\nobreak\thinspace}
\DeclareUnicodeCharacter{200B}{\hspace{0pt}}

\newcommand{\uvec}{\mathbf{u}}
\newcommand{\vvec}{\mathbf{v}}
\newcommand{\Lvec}{\mathbf{L}}
\newcommand{\Lphi}{\mathbf{L}_\Phi}
\newcommand{\LA}{\mathbf{L}_A}

\newcommand{\AL}{\mathbf{A}_L}

\newcommand{\Wvec}{\mathbf{W}}
\newcommand{\Bvec}{\mathbf{B}}
\newcommand{\Evec}{\mathbf{E}}
\newcommand{\Rvec}{\mathbf{R}}
\newcommand{\Jvec}{\mathbf{J}}
\newcommand{\Dvec}{\mathbf{D}}
\newcommand{\omvec}{\boldsymbol{\omega}}
\newcommand{\tvec}{\boldsymbol{\theta}}
\newcommand{\phivec}{\boldsymbol{\phi}}
\newcommand{\rvec}{\mathbf{r}}
\newcommand{\xvec}{\mathbf{x}}
\newcommand{\yvec}{\mathbf{y}}
\newcommand{\zvec}{\mathbf{z}}
\newcommand{\kvec}{\mathbf{k}}
\newcommand{\pvec}{\mathbf{p}}
\newcommand{\qvec}{\mathbf{q}}
\newcommand{\jvec}{\mathbf{j}}

\newcommand{\lvec}{\mathbf{l}}
\newcommand{\avec}{\mathbf{a}}
\newcommand{\bvec}{\mathbf{b}}

\title{An $SO(3)$ Gauge Theory of Turbulence with Spontaneous Symmetry Breaking}
\author{Ahmed Farooq}
\date{}

\begin{document}

\maketitle

\begin{abstract}
Fully developed isotropic turbulence exhibits a dual nature: a continuous, scale‑invariant energy cascade coexists with discrete, intense vortex filaments.  We show that this duality arises from a spontaneously broken $SO(3)$ gauge symmetry.  By identifying the specific angular momentum $\Lvec = \rvec\times\uvec$ as a non‑Abelian gauge connection and the radial velocity $u_r$ as a Higgs field, the turbulent vacuum is described by the $SO(3)$ Georgi–Glashow model.  When the radial strain condenses, the symmetry breaks $SO(3)\to U(1)$, generating a topological mass gap $M_W = gv$.  This gap partitions the energy into a massless $U(1)$ sector (the solenoidal background) that sustains the Kolmogorov cascade, and a massive $SO(3)/U(1)$ sector that is confined to vortex filaments.

Using high‑resolution DNS data (JHTDB, $Re_\lambda\approx433$), we empirically verify three key predictions:  
(i) the energy spectra obey a strict $1:2$ equipartition over the inertial range, with a sharp divergence at $M_W \approx 40$;  
(ii) The radial Higgs field extracted around isolated vortex cores follows the exact BPS monopole profile $H(r)=\coth(r/\eta)-\eta/r$ with $\eta = 0.0093$ domain units and the VEV $v = 0.338$, identifying the ubiquitous ``worms'' as macroscopic 't Hooft–Polyakov monopoles.
(iii) the Wilson loop computed from the velocity field exhibits a clean area law $\langle W_C \rangle \sim e^{-\sigma A}$ with string tension $\sigma = 0.303 \pm 0.009$, directly confirming the confining nature of the turbulent vacuum.

Motivated by these observations, we construct an explicit cylindrical 't Hooft–Polyakov string defect – the non‑Abelian analog of the Abrikosov vortex – whose transverse profile is exactly the same BPS function.  The mass gap is tied to the Kolmogorov scale via $\eta\approx 3.2\,\eta_K$, directly linking kinematic viscosity to the Higgs mechanism.  The dilute gas of strings implies the Wilson area law, characterizing the turbulent vacuum as a confining phase.  Our results provide the first empirical evidence of spontaneous symmetry breaking and topological monopoles in a classical fluid system, and establish a direct connection between turbulence and non‑Abelian gauge theory.
\end{abstract}

\section{Introduction}

Fully developed isotropic turbulence presents a dual nature: a continuous, scale‑invariant energy cascade coexists with the spontaneous emergence of discrete, intense vortex filaments – the ``worms'' first observed numerically by Vincent and Meneguzzi~\cite{VincentMeneguzzi1991} and later characterized by Jiménez et al.~\cite{Jimenez1993}, Ishihara et al.~\cite{Ishihara2009} and others. While Kolmogorov's 1941 theory successfully predicts the $k^{-5/3}$ energy spectrum and the $4/5$ law, it does not explain the origin of these coherent structures nor their relation to the statistical cascade. A unified first‑principles description has remained elusive.

\subsection*{Geometric and gauge approaches to turbulence}

The deep geometric structure of fluid flow was recognised early by Arnold~\cite{Arnold1966}, who showed that the Euler equations describe geodesic flow on the group of volume‑preserving diffeomorphisms. This insight led to Hamiltonian reduction theories~\cite{marsden1983reduction} and the Euler–Poincaré formulation~\cite{holm1998euler}. Independently, Moffatt~\cite{moffatt1969degree} introduced topological concepts through helicity and knotted vortex tubes, and Freedman and He~\cite{freedman1991non} further advanced topological concepts in ideal fluid mechanics by developing a lower bound for the kinetic energy.\\

In the 1990s, Polyakov~\cite{polyakov1995two} applied two‑dimensional conformal field theory to turbulence, while Migdal~\cite{migdal1995loop} developed a loop equation formalism treating turbulence as a gauge theory with Wilson loops. Direct analogies between fluid dynamics and Yang–Mills theory appeared in the works of Holm and Kupershmidt~\cite{HolmKupershmidt1983}, Jackiw, Pi, and Polychronakos~\cite{Jackiw2002}, and Kambe~\cite{Kambe2007}. Table~\ref{tab:gauge-fluid} (Appendix C) summarises these historical formulations. Despite these advances, a predictive gauge theory yielding verifiable empirical signatures – such as a sharp mass gap or a topological defect profile – has been missing.

\subsection*{Renormalization group approaches}

A parallel line of inquiry has focused on calculating the statistical properties of turbulence from first principles using renormalization group (RG) methods.  The modern field-theoretic approach to turbulence was pioneered by Forster, Nelson, and Stephen \cite{ForsterNelsonStephen1977}, who applied the RG to the randomly stirred Navier–Stokes equation.  Subsequently, Yakhot and Orszag \cite{YakhotOrszag1986} developed a dynamical RG procedure and obtained a value for the Kolmogorov constant $C_K \approx 1.6$, in good agreement with experiment and DNS.  This work has been critically reassessed by Eyink \cite{Eyink1994}, who questioned some of its formal foundations.  Independently, McComb and Watt \cite{McCombWatt1992} developed a two-field theory of incompressible turbulence that also yields a Kolmogorov constant in this range.  For a comprehensive overview, see the recent review by Verma \cite{Verma2025}.\\

While these RG approaches provide a theoretical basis for the inertial-range spectrum, they do not explain the coexistence of the continuous cascade with discrete vortex filaments – the duality that motivates our gauge-theoretic formulation.

\subsection*{Angular momentum as a gauge field}

In companion works~\cite{FarooqBeyond2026,FarooqIsotropic2026}, we introduced an angular‑momentum based reformulation of fluid mechanics. The specific angular momentum field $\Lvec = \rvec\times\uvec$ (with $\rvec$ a fixed separation vector) obeys a transport equation and separates into coherent and incoherent parts via a Helmholtz decomposition:

\begin{equation}
\Lvec = -\nabla\Phi_L + \nabla\times\AL \equiv \Lphi + \LA
\end{equation}

The corresponding velocity fields $\uvec_\Phi$ and $\uvec_A$ satisfy a universal energy partition $E_\Phi:E_A=1:2$ $E_\Phi = \frac{1}{2} \int |\Lvec_\Phi|^2 \;dV$ and $E_A = \frac{1}{2} \int |\Lvec_A|^2 \;dV$  verified by DNS. The radial component $\uvec_r = \uvec - \uvec_\Phi - \uvec_A$ behaves as a scalar order parameter. This structure is strongly reminiscent of a spontaneously broken gauge theory.

\subsection*{From kinematics to the Georgi–Glashow model}

Under a local $SO(3)$ rotation $R(\xvec)$, the ordinary derivative of $\uvec$ does not transform covariantly. To restore covariance one must introduce a gauge connection. We identify this connection with the scaled specific angular momentum:

\begin{equation}
\Wvec_\mu = \frac{1}{\eta^2}\,\Lvec_\mu,\qquad g = \eta^{-1}
\end{equation}

where $\eta$ is a topological microscale. \\

The radial velocity becomes a Higgs field in the adjoint representation:

\begin{equation}
\boldsymbol{\phi} = \frac{\uvec_r}{U_0}
\end{equation}

with $U_0$ the rms velocity. The simplest $SO(3)$‑invariant Lagrangian containing both a gauge field and an adjoint scalar is the Georgi–Glashow model (the $SO(3)$ Yang–Mills–Higgs theory):

\begin{equation}
\mathcal{L} = -\frac14 F_{\mu\nu}^a F^{a\mu\nu} + \frac12 (D_\mu\boldsymbol{\phi})^a (D^\mu\boldsymbol{\phi})^a - \frac{\lambda}{4}\bigl(|\boldsymbol{\phi}|^2 - v^2\bigr)^2
\end{equation}

The fields $\Wvec_\mu$ and $\phivec$ that appear in the Lagrangian are not the instantaneous velocity and strain fields of the Navier–Stokes equations. Rather, they are coarse‑grained collective variables that encode the statistical properties of the turbulent ensemble—moments, correlations, and scale interactions—in the neighbourhood of each point. The Lagrangian $\mathcal{L}$ is an effective action for these statistical degrees of freedom, analogous to the Landau–Ginzburg action for a critical system.  The gauge symmetry $SO(3)$ is a symmetry of this effective description, not of the microscopic Navier–Stokes equations.  This is the standard interpretation in statistical field theory and is essential for understanding the role of spontaneous symmetry breaking and topological defects in the turbulent vacuum.\\

With the above motivating summary in place, we proceed to systematically develop the theory.\\

The paper is organized as follows. Section~2 introduces the $\Lvec$ framework and the decomposition into radial and tangential parts. Section~3 presents the Yang–Mills Lagrangian and its relation to turbulence observables. Section~4 develops the Higgs sector and the spontaneous symmetry breaking, leading to the Georgi–Glashow model. Section~5 develops the 't Hooft Polyakov monopole theory and Section~6 proposes a new topological object, the 't Hooft Polyakov thread.  Section~7 develops the energy partition in the Hamiltonian recovering the results of \cite{FarooqIsotropic2026}.  Section~8 provides the dictionary between gauge fields and fluid variables. Section~9 starts the validations with the recovery of Zakharov's weak turbulence theory.  Section 10 validates the theory against the $1:2$ spectral partition and the mass gap. Section~11 demonstrates that the vortex filaments are ’t Hooft–Polyakov monopoles via the BPS profile fit. Section~12  shows that the Wilson Loops lead to the Kelvin circulation theorem. Finally we have the conclusions and discussion of future outlook section.

\newpage

\part{Formalism}
\section{The $\Lvec$ Framework and $SO(3)$ Invariance}
\label{sec:Lvec-framework}

In this section we summarize the angular-momentum based reformulation of isotropic turbulence developed in companion papers \cite{FarooqBeyond2026,FarooqIsotropic2026}. This framework provides the geometric foundation upon which the gauge theory is built.

\subsection{Angular momentum transport and Helmholtz decomposition}

Define the specific angular momentum field relative to a fixed separation vector $\rvec$:

\begin{equation}
\Lvec(\xvec,t) = \rvec \times \uvec(\xvec,t)
\end{equation}

Its transport equation follows from the Navier–Stokes equations:

\begin{equation}
\frac{\partial \Lvec}{\partial t} + \uvec \cdot \nabla \Lvec = -\frac{1}{\rho} \nabla p + \nu (\nabla^2 \Lvec - 2\omvec)
\label{eq:Ltransport}
\end{equation}

where $\omvec = \nabla\times\uvec$ is the vorticity. The curl and divergence of $\Lvec$ are given by:

\begin{align}
\nabla \times \Lvec &= -2\uvec - (\rvec\cdot\nabla)\uvec \label{eq:curlL}\\
\nabla \cdot \Lvec &= \rvec \cdot \omvec \label{eq:divL}
\end{align}

The Helmholtz decomposition separates $\Lvec$ into a curl‑free (longitudinal) part and a divergence‑free (transverse) part termed as the coherent and background fields (\cite{FarooqIsotropic2026}) :

\begin{equation}
\Lvec = \Lphi + \LA; \qquad 
\Lphi = -\nabla\Phi_L ;\quad 
\LA = \nabla\times\AL;
\label{eq:Ldecomp}
\end{equation}

a scalar potential and $\AL$ is a vector potential. The generalized energy of the angular momentum field, the ``edicity'' is defined as:

\begin{align}
E_\phi &= \int |\Lvec_\Phi|^2 dV\\
E_A &= \int |\LA|^2dV
\label{eqn:editicity-def}
\end{align}

The corresponding velocity fields are reconstructed as:

\begin{align}
\uvec_\Phi &= \frac{1}{r^2}\,\Lphi\times\rvec \label{eq:uPhi}\\
\uvec_A &= \frac{1}{r^2}\,\LA\times\rvec \label{eq:uA}
\end{align}

The total velocity then decomposes into three orthogonal components:

\begin{equation}
\uvec = \uvec_\Phi + \uvec_A + \uvec_r
\label{eq:totalvelocity}
\end{equation}

where $\uvec_r = u_r\,\hat{\rvec}$ is the radial (dilatational) part. The radial field is defined implicitly as the part of $\uvec$ that does not contribute to $\Lvec$; indeed $\rvec\times\uvec_r = 0$. This decomposition is exact and kinematic: it follows solely from the definition of $\Lvec$ and the Helmholtz theorem. Taking the curl of Eq.~\ref{eq:totalvelocity}, we obtain the cooresponding decomposition of the vorticity field:

\begin{equation}
\omvec = \omvec_\Phi + \omvec_A + \omvec_r
\label{eq:voriticity-decomposition}
\end{equation}

In \cite{FarooqIsotropic2026} a statistical mechanical analysis of the decomposition (\ref{eq:totalvelocity}) predicted a universal energy partition among the three components. In the inertial range, the kinetic energies satisfy:

\begin{equation}
E_{\Phi}^u : E_{A}^u : E_{r}^u = 1 : 2 : 3
\label{eq:partition-ratio}
\end{equation}

where $E_{\Phi}^u = \frac12\langle|\uvec_\Phi|^2\rangle$, $E_{A}^u = \frac12\langle|\uvec_A|^2\rangle$, and $E_{r}^u = \frac12\langle|\uvec_r|^2\rangle$. Equivalently, the tangential part ($\uvec_\tau = \uvec_\Phi + \uvec_A$) and the radial part obey $E_r^u : E_\tau^u = 1:3$, while the tangential part further splits as $E_\Phi^u : E_A^u = 1:2$. These ratios were verified against high‑resolution DNS data, as shown in \cite{FarooqIsotropic2026}.\\

Several identities follow from the definitions. Using (\ref{eq:curlL}) and (\ref{eq:Ldecomp}) we have:

\begin{equation}
\nabla\times\LA = \nabla\times\Lvec = -2\uvec - (\rvec\cdot\nabla)\uvec
\label{eq:curlLA}
\end{equation}

and consequently

\begin{equation}
\nabla^2\AL = 2\uvec + (\rvec\cdot\nabla)\uvec
\label{eq:LaplaceA}
\end{equation}

Moreover, from (\ref{eq:divL}) and $\Lphi = -\nabla\Phi_L$ we obtain:

\begin{equation}
\nabla^2\Phi_L = -\nabla\cdot\Lphi  = -\rvec\cdot\omvec 
\label{eq:PhiL}
\end{equation}

since $\nabla\cdot\LA=0$ in the coulomb gauge. These relations will be essential when we map gauge‑theoretic quantities to fluid observables.

\subsection{From global to local $SO(3)$ symmetry}

The decomposition (\ref{eq:totalvelocity}) separates the flow into a radial component $\uvec_r$ (which does not contribute to angular momentum) and a tangential component $\uvec_\tau$ (which does). Under a global $SO(3)$ rotation $\uvec \to R\uvec$, the angular momentum transforms as $\Lvec \to R\Lvec$, and the separation into radial and tangential parts is preserved. However, in a turbulent flow the rotational structures (eddies, vortices) can have different orientations at different points. It is therefore natural to consider local $SO(3)$ transformations $R(\xvec)$. Under a local $SO(3)$ rotation $R(\xvec)$ of the coarse‑grained velocity field $\langle\uvec\rangle$, the ordinary derivative does not transform covariantly.  This suggests that a gauge connection must be introduced for the effective description.  (The microscopic Navier–Stokes equations, of course, are not locally $SO(3)$ invariant; the gauge symmetry is a property of the coarse‑grained effective theory.)\\

We propose to identify this gauge connection with the scaled angular momentum:

\begin{equation}
\Wvec_\mu \equiv \frac{1}{\eta^2}\,\Lvec_\mu
\label{eq:Wdef}
\end{equation}

where $\eta$ is a characteristic microscale (later identified with the vortex core radius). We also introduce the temporal part of the gauge connection:

\begin{equation}
\Wvec \equiv (-H, \Lvec)
\end{equation}

where $H$ is a scalar which we will identify with the Bernoulli head:  $-H = \frac{p}{\rho} + \frac{u^2}{2}$. We thus have $W_0 = H$ and $W_i = L_i$, with $i=1,2,3$.  In the temporal gauge we set $W_0=0$.\\

The radial velocity provides a natural Higgs field:

\begin{equation}
\phivec \equiv \frac{\uvec_r}{U_0}
\label{eq:phidef}
\end{equation}

with $U_0$ the rms velocity. In Appendix B we construct a gauge‑covariant shifted velocity $\tilde{\uvec} = \uvec + \Wvec\times\rvec$ and show that after coarse‑graining it transforms covariantly. The Lagrangian will then be built from $\Wvec$ and $\phivec$ in a locally $SO(3)$ invariant manner, leading naturally to the Georgi–Glashow model.\\

The empirical success of the $1:2:3$  partition (Eq.~\ref{eq:partition-ratio}), together with the geometric structure of the $\Lvec$ decomposition, strongly suggests that an underlying gauge principle is at work. The remainder of this paper is devoted to making this principle explicit and testing its most striking predictions.\\

\begin{figure}[htbp]
\centering
\includegraphics[width=0.7\linewidth]{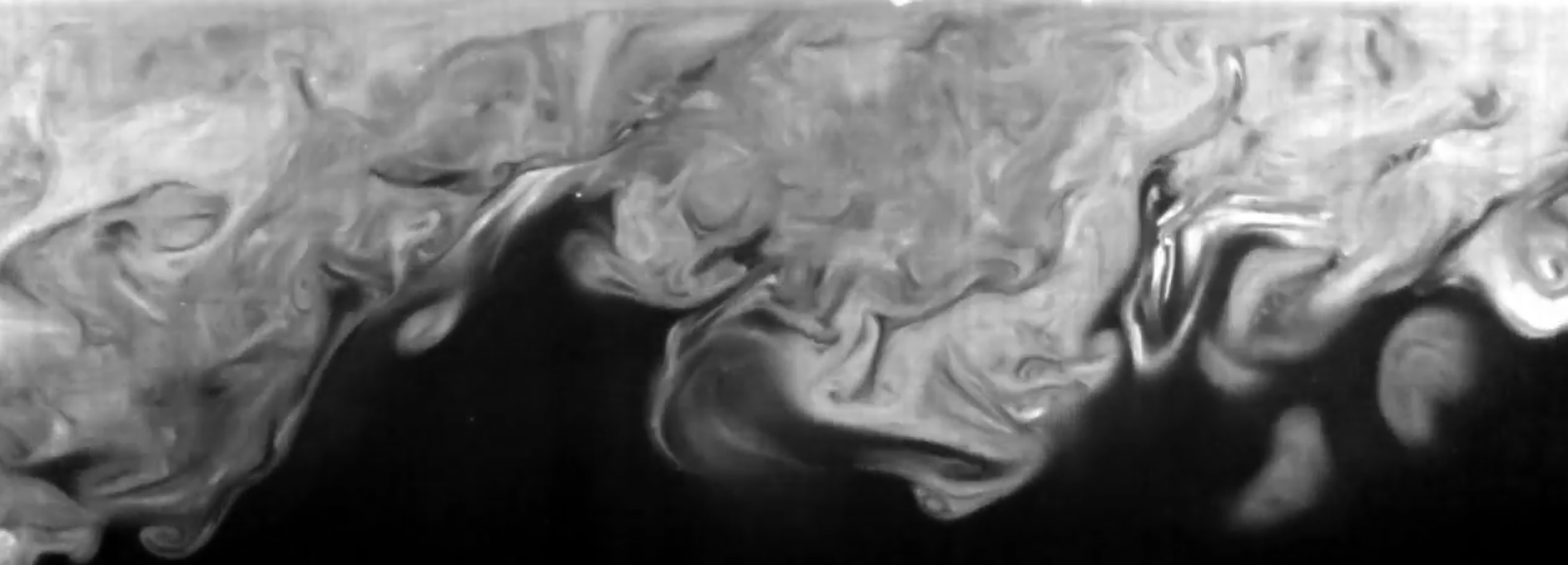}
\caption{Turbulent Boundary Layer (\cite{lee2012spatiallydevelopingturbulentboundary}) showing eddies of different scales at different locations.  The independent orientation and rotation rate of each eddy motivate the need for a local $SO(3)$ gauge symmetry.}
\label{fig:vonkarman}
\end{figure}

The transformation properties of $\Lvec$ under local rotations point directly to a gauge structure. Under a global $SO(3)$ rotation $\uvec \to R\uvec$ the Navier–Stokes equations are invariant. However, turbulent flows exhibit local rotational structures (eddies, vortices) that can rotate independently at different points (see Fig.~\ref{fig:vonkarman}). If we demand that the effective coarse‑grained description be invariant under local rotations $R(\xvec)$, the ordinary derivative fails because:

\begin{equation}
\partial_i (R\uvec) = (\partial_i R)\uvec + R\,\partial_i\uvec \neq R\,\partial_i\uvec 
\end{equation}

\begin{figure}[htbp]
\centering
\begin{tikzpicture}[scale=0.6]
\begin{scope}
\draw[gray, very thin] (0,0) grid (6,6);
\foreach \x in {0,1,...,6} 
    \foreach \y in {0,1,...,6} 
        \node at (\x,\y) [circle, inner sep=0.5pt, fill=black] {};

\foreach \x in {1,2,...,5} 
    \foreach \y in {1,2,...,5} 
        \draw[blue, ->] (\x,\y) -- (\x+0.3,\y);
        
\node[above] at (3,6.5) {Original field $\uvec(\xvec)$};
\end{scope}

\begin{scope}[shift={(9,0)}]
\draw[gray, very thin] (0,0) grid (6,6);
\foreach \x in {0,1,...,6} 
    \foreach \y in {0,1,...,6} 
        \node at (\x,\y) [circle, inner sep=0.5pt, fill=black] {};

\foreach \x in {1,2,...,5} 
    \foreach \y in {1,2,...,5} {
        \pgfmathsetmacro{\angle}{20*(\x-3)+20*(\y-3)}
        \draw[red, ->, rotate around={\angle:(\x,\y)}] (\x,\y) -- (\x+0.3,\y);
    }
        
\node[above] at (3,6.5) {Locally rotated $R(\xvec)\uvec(\xvec)$};
\end{scope}

\draw[->, thick] (6.5,3) -- node[above] {Local rotation $R(\xvec)$} (8.5,3);

\node[below, align=center] at (3,-0.5) {Global: one $R$\\for entire flow};
\node[below, align=center] at (12,-0.5) {Local: different $R(\xvec)$\\at each point};
\end{tikzpicture}
\caption{Illustration of global versus local $SO(3)$ symmetry.  Left: the original velocity field with a uniform orientation.  Right: the field after applying a spatially varying rotation $R(\xvec)$, creating a vortex‑like pattern.  A global rotation (one $R$ for the entire flow) preserves the Navier–Stokes equations, but a local rotation (different $R$ at each point) breaks covariance; restoring it requires a gauge connection $\Wvec_\mu$.}
\label{fig:global-local-symmetry}
\end{figure}
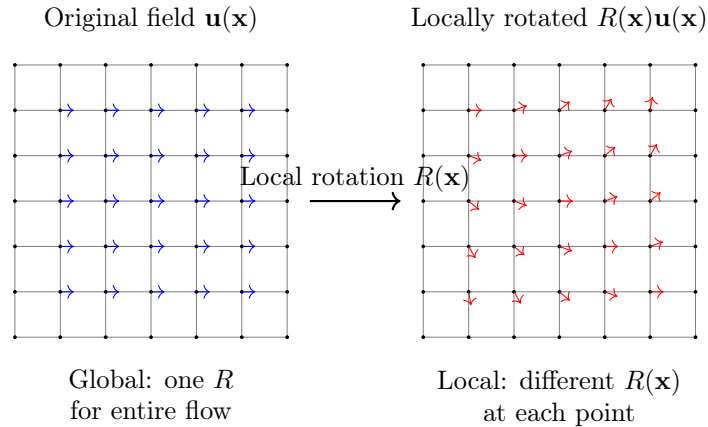

In gauge theory this is repaired by introducing a connection $\Wvec_i$ and replacing $\partial_i$ with the covariant derivative $D_i = \partial_i + g\Wvec_i\times$. The connection must transform as $\Wvec_i \to R\Wvec_i R^{-1} + (\partial_i R)R^{-1}$. Fig.~\ref{fig:global-local-symmetry} shows the difference between global and local $SO(3)$ symmetry and illustrates why the latter may be more suitable for capturing the dynamics of turbulent fields.\\

Having identified the gauge connection (Eq.~\ref{eq:Wdef}) and the Higgs field (Eq. \ref{eq:phidef}), a gauge‑covariant shifted velocity is then defined as:

\begin{equation}
\tilde{\uvec} = \uvec + \Wvec \times \rvec 
\end{equation}

Since $\Wvec$ has dimensions $[T^{-1}]$ and $\rvec$ dimensions $[L]$, $\tilde{\uvec}$ has velocity dimensions. Under a local $SO(3)$ transformation, a direct calculation (see Appendix B) shows that the non‑covariant terms are proportional to $\rvec$ or its scalar products. In a statistically homogeneous and isotropic turbulent field, ensemble averaging over the random separation $\rvec$ causes these terms to vanish, yielding:

\begin{equation}
\langle \tilde{\uvec} \rangle \;\longrightarrow\; R\,\langle \tilde{\uvec} \rangle 
\end{equation}

Thus, after coarse‑graining, $\tilde{\uvec}$ behaves as a covariant field. This justifies the construction of a Yang–Mills–Higgs Lagrangian for the coarse‑grained turbulent variables.

The image in Fig.~\ref{fig:vonkarman} depicts a turbulent boundary layer (Monty et al. \cite{lee2012spatiallydevelopingturbulentboundary}) which vividly illustrate the phenomenon of local, scale‑dependent rotational structures.  Eddies of different sizes appear at different locations, each rotating at its own rate and orientation.  A purely global $SO(3)$ symmetry (a single rotation applied everywhere) cannot capture this richness: a large eddy in one region may be rotating clockwise while a small eddy nearby rotates counter‑clockwise.  To respect the independence of these local rotations, the symmetry must be promoted to a \emph{local} $SO(3)$ gauge symmetry, where the rotation angle varies from point to point.  This naturally introduces a gauge connection $\Wvec_\mu$ that compensates for the inhomogeneous rotations and leads to the Yang–Mills structure developed in this paper.

\subsection{Angular Momentum as an $SO(3)$ Gauge Connection}
\label{sec:gauge_connection}

In standard continuum mechanics, the local angular momentum field of a fluid, defined as $\Lvec(\rvec) = \rvec \times \uvec(\xvec)$, is conventionally treated as a coordinate-dependent quantity. However, when formulating a local gauge theory of isotropic turbulence, the inherent origin-dependence of $\Lvec$ reveals a non-trivial mathematical structure: it behaves not as a covariant tensor, but precisely as a non-Abelian gauge connection.\\

To see this, recall the behavior of a generic gauge field $A_\mu$ in a Yang-Mills theory. Under a local gauge transformation $U(x)$, the connection transforms via a shift:

\begin{equation}
A_\mu \to U A_\mu U^\dagger + \frac{i}{g} U \partial_\mu U^\dagger
\label{eq:standard_gauge_transform}
\end{equation}

The presence of the inhomogeneous second term dictates that $A_\mu$ cannot be a direct physical observable, as its local magnitude is entirely dependent on the arbitrary choice of gauge.\\

We observe an identical behavior in the fluid's angular momentum field. Let us apply a spatial translation to the system, shifting the reference frame origin by an arbitrary constant vector $\rvec_0$. The position vector transforms as $\rvec \to \rvec' = \rvec - \rvec_0$. Consequently, the local angular momentum field transforms as:

\begin{equation}
\Lvec'(\rvec) = (\rvec - \rvec_0) \times \uvec(\rvec) = \Lvec(\rvec) - \rvec_0 \times \uvec(\rvec)
\label{eq:L_transform}
\end{equation}

Equation \ref{eq:L_transform} shows that $\Lvec$ does not transform trivially under translation. Instead, it accumulates an inhomogeneous shift, $-\rvec_0 \times \uvec(\rvec)$.   This spatial translation mirrors the structural form of the gauge connection transformation (in Eq.~\ref{eq:standard_gauge_transform}), where the field acquires an additive term under a symmetry operation. (We note that the inhomogeneous term here is proportional to the local velocity, rather than the derivative of a gauge parameter; this reflects the kinematic origin of the angular momentum field, but the structural parallel is sufficient to motivate the gauge connection interpretation).\\

By promoting the global rotational symmetry of the fluid to a local $SO(3)$ gauge symmetry, the freedom to arbitrarily place the coordinate origin becomes the formal gauge freedom of the system. The local magnitude of $\Lvec$ is, therefore, a gauge-dependent artifact rather than an objective physical density.\\

Consequently, extracting true topological or dynamical invariants from the turbulent cascade requires handling this origin-dependence mathematically as a gauge redundancy. In lattice gauge theory, physical observables are isolated by integrating over all possible gauge configurations (the gauge volume). Similarly, in an isotropic and homogeneous turbulent field, no single coordinate origin possesses privileged physical status. Evaluating the angular momentum field by averaging over an ensemble of spatial origins acts as a fluid-dynamical analog to gauge integration. This procedure systematically factors out the coordinate-induced shifts, isolating the true, origin-independent topological structures of the turbulent cascade.  We will utilize this procedure in Part II of this article where we perform computations on DNS data.

\section{Yang–Mills Lagrangian and Field Strength Tensor}
\label{sec:field-strength-lorentzian}

We work with the gauge group $SO(3)$.  The Lie algebra, its generators, and the equivalence between the vector (cross‑product) and matrix (commutator) formulations are summarized in Appendix~A.  Following the standard language of quantum chromodynamics (QCD), we refer to the index $a = 1,2,3$ as a \textit{color index}.  The gauge field $\Wvec_\mu = (W_\mu^1, W_\mu^2, W_\mu^3)$ thus carries color, and the non‑Abelian field strength $F_{\mu\nu}^a$ contains cubic and quartic self‑interactions that are the analogs of gluon interactions in QCD.\footnote{In analogy with QCD, where each generator $T^a$ ($a=1,\dots,8$) is associated with a gluon field $G_\mu^a$, in our $SO(3)$ gauge theory the three generators $T^1,T^2,T^3$ are associated with the gauge bosons $W_\mu^1, W_\mu^2, W_\mu^3$.  In the broken phase, $W_\mu^1$ and $W_\mu^2$ acquire masses $M_W$ (the broken generators), while $W_\mu^3$ remains massless and corresponds to the unbroken $U(1)$ sector.}\\

With the gauge connection $\Wvec_\mu$ identified as the coarse‑grained specific angular momentum $\Wvec_\mu = \eta^{-2}\langle\Lvec_\mu\rangle$, we now construct the kinetic term for the gauge field.  (Here and throughout, angular brackets denote coarse‑graining over a scale $\ell \gg \eta_K$.) We work in the temporal gauge $W_0^a = 0$, which eliminates unphysical degrees of freedom and simplifies the interpretation.  The covariant derivative of the shifted velocity field is:

\begin{equation}
D_\mu \tilde{\uvec} = \partial_\mu \tilde{\uvec} + \Wvec_\mu \times \tilde{\uvec}
\end{equation}

and the field strength (curvature) of the connection is defined as:

\begin{equation}
F_{\mu\nu}^a = \partial_\mu W_\nu^a - \partial_\nu W_\mu^a + g \epsilon^{abc} W_\mu^b W_\nu^c;
\qquad 
\mathbf{F}_{\mu\nu} = F_{\mu\nu}^a T^a
\end{equation}

The coupling constant $g$ has dimensions of inverse length and is related to the topological microscale by $g = \eta^{-1}$.  In a turbulent flow, $\eta$ sets the size of the defect core; as $g$ increases the core shrinks, concentrating the non‑Abelian magnetic flux.\\

The gauge‑invariant kinetic term for the connection is the Yang–Mills Lagrangian:

\begin{equation}
\mathcal{L}_{\mathrm{YM}} = -\frac14 F_{\mu\nu}^a F^{a\mu\nu}
\label{eq:YM-lagrangian}
\end{equation}

where the Minkowski metric $\eta_{\mu\nu}=\mathrm{diag}(+1,-1,-1,-1)$ is used.  Expanding the square yields three types of contributions:

\begin{align}
\mathcal{L}_2 &= (\partial_\mu W_\nu^a - \partial_\nu W_\mu^a)(\partial^\mu W^{a\nu} - \partial^\nu W^{a\mu}) \\
\mathcal{L}_3 &= 2g \,\epsilon^{abc} (\partial_\mu W_\nu^a) W^{b\mu} W^{c\nu} \\
\mathcal{L}_4 &= g^2 \,\epsilon^{abc}\epsilon^{ade} W_\mu^b W_\nu^c W^{\mu d} W^{\nu e}
\end{align}

The total Lagrangian density is $\mathcal{L}_{\mathrm{YM}} = -\frac14(\mathcal{L}_2 + \mathcal{L}_3 + \mathcal{L}_4)$.

\subsection{Physical interpretation in turbulence}

The quadratic term $\mathcal{L}_2$ describes the propagation of free waves.  Using the Helmholtz decomposition $\Wvec \propto -\nabla\Phi_L + \nabla\times\AL$, the spatial part of $\mathcal{L}_2$ becomes:

\begin{equation}
\mathcal{L}_2 \supset \frac12 \bigl[ (\partial_0\Wvec)^2 - (\nabla\times\Wvec)^2 \bigr]
\end{equation}

Since $\nabla\times\nabla\Phi_L = 0$, the curl operator isolates the solenoidal component $\nabla\times\AL$ – the massless $U(1)$ sector.  The longitudinal part $\nabla\Phi_L$ does not appear in the magnetic term; it will acquire a mass through the Higgs mechanism.  Thus, even at the level of the free theory, the quadratic Lagrangian already encodes the geometric separation of the turbulent velocity into a massless background and a massive coherent part.\\

The cubic term $\mathcal{L}_3$ is responsible for three‑wave interactions.  In the fluid interpretation it corresponds to vortex stretching and the nonlinear transfer of energy across scales.  Indeed, using the relation

\begin{equation}
\nabla\times\Wvec = -2\uvec - (\rvec\cdot\nabla)\uvec
\label{eq:W-curl}
\end{equation}

(derived in the $\Lvec$ framework), one finds that $\mathcal{L}_3$ generates terms of the form $\uvec\cdot(\Wvec\times\Wvec)$ that describe the interaction of the gauge field with itself – the non‑Abelian analog of the Lamb vector.\\

The quartic term $\mathcal{L}_4$ gives four‑point interactions, which are essential for the renormalization of the coupling and for producing non‑Gaussian statistics.  In turbulence, these terms are responsible for small‑scale intermittency and the formation of heavy tails in the probability distribution of velocity gradients.\\

Thus, the $SO(3)$ Yang–Mills Lagrangian provides a geometric, first‑principles description of the rotational (tangential) dynamics of turbulence.  In the next section we introduce the Higgs sector to account for the radial (dilatational) motions, completing the Georgi–Glashow model.

\subsection{Color‑electric and color‑magnetic fields}

We define the electric and magnetic fields, using the standard definitions (in the metric $\eta_{\mu\nu}=\operatorname{diag}(1,-1,-1,-1)$):

\begin{align}
E_i^a &= F_{0i}^a = \partial_0 W_i^a - \partial_i W_0^a + g\,\epsilon^{abc}W_0^b W_i^c,\\
B_i^a &= \frac12\epsilon_{ijk}F_{jk}^a = \epsilon_{ijk}\Bigl(\partial_j W_k^a + \frac{g}{2}\epsilon^{abc}W_j^b W_k^c\Bigr)  \label{eq:color-magnetic}
\end{align}

These are the colour‑electric and colour‑magnetic fields.\\

It may be noted that in the temporal gauge $W_0^a=0$, the field strength components split into electric field becomes $E_i^a = F_{0i}^a = \partial_0 W_i^a $.  We can write the field strength  tensor (in both covariant and contravariant forms) explicitly in matrix notation as:

\begin{equation}
F_{\mu\nu}^a = \begin{pmatrix}
0 & -E_1^a & -E_2^a & -E_3^a \\[4pt]
E_1^a & 0 & B_3^a & -B_2^a \\[4pt]
E_2^a & -B_3^a & 0 & B_1^a \\[4pt]
E_3^a & B_2^a & -B_1^a & 0
\end{pmatrix}; \quad  \quad F^{a\mu\nu} = 
\begin{pmatrix}
0 & E_1^a & E_2^a & E_3^a \\[4pt]
-E_1^a & 0 & B_3^a & -B_2^a \\[4pt]
-E_2^a & -B_3^a & 0 & B_1^a \\[4pt]
-E_3^a & B_2^a & -B_1^a & 0
\end{pmatrix}
\end{equation}

In compact vector notation (suppressing the color index $a$),

\begin{align}
\Evec &= \partial_0\Wvec\\
\Bvec &= \nabla\times\Wvec + \frac{g}{2}\,\Wvec\times\Wvec \label{color-magnetic-vector}
\end{align}

where the cross product in the magnetic term acts simultaneously on spatial and color indices via $(\Wvec\times\Wvec)^a_i = \epsilon^{abc}\epsilon_{ijk}W_j^bW_k^c$.  The Yang–Mills Lagrangian then takes the Maxwell‑like form:

\begin{equation}
\mathcal{L}_{\mathrm{YM}} = \frac12\bigl(\Evec\cdot\Evec - \Bvec\cdot\Bvec\bigr)
\end{equation}

The non‑Abelian nature is hidden in the additional $(\Wvec\times\Wvec)$ term of $\Bvec$, which gives rise to cubic and quartic self‑interactions.\\

The connection between the gauge fields and fluid variables is established through the kinematic relation derived from the $\Lvec$ framework (Eq.~\ref{eq:curlL}):

\begin{equation}
\nabla \times \Wvec = -2\uvec - (\rvec\cdot\nabla)\uvec 
\end{equation}

Using this identity, the quadratic part of the Lagrangian becomes:

\begin{equation}
\mathcal{L}_2 = \frac12 (\partial_0\Wvec)^2 - 2u^2 - 2\,\uvec\cdot(\rvec\cdot\nabla\uvec) - \frac12 (\rvec\cdot\nabla\uvec)^2 
\end{equation}

Each term can be interpreted in turbulence language after ensemble averaging over an isotropic field: $\frac12 (\partial_0\Wvec)^2$ is the time‑derivative contribution (related to the unsteady part of the angular momentum). $-2u^2$ is proportional to the negative of the turbulent kinetic energy density $-\frac12 \langle u^2\rangle$.\\

The term $-\uvec\cdot(\rvec\cdot\nabla\uvec)$ mixes velocity and velocity gradient.  Decomposing $\nabla\uvec$ into symmetric (strain) and antisymmetric (vorticity) parts, the antisymmetric part gives a measure of helicity $\langle \uvec\cdot\omvec\rangle$ after isotropic averaging. $-\frac12 (\rvec\cdot\nabla\uvec)^2$ involves quadratic products of velocity gradients; its trace under isotropic averaging is proportional to the enstrophy density $\frac12\langle\omvec^2\rangle$.\\

Thus $\mathcal{L}_2$ encodes the kinetic energy, helicity, and enstrophy of the turbulent flow.\\

The cubic term:

\begin{equation}
\mathcal{L}_3 = -g\,(\nabla\times\Wvec)\cdot(\Wvec\times\Wvec)
\end{equation}

using Eq.~\ref{eq:curlL} becomes

\begin{equation}
\mathcal{L}_3 = -g\bigl(-2\uvec - \rvec\cdot\nabla\uvec\bigr)\cdot(\Wvec\times\Wvec)
\end{equation}

The quantity $\Wvec\times\Wvec$ represents the non‑Abelian part of the magnetic field.  This term generates three‑wave interactions and is responsible for vortex stretching and the nonlinear energy transfer across scales – the core of the turbulent cascade (which we relate to the Zakharov wave turbulence in Part II of the paper).\\

The quartic term,

\begin{equation}
\mathcal{L}_4 = -\frac{g^2}{8}\,(\Wvec\times\Wvec)^2
\end{equation}

comes from the square of the non‑Abelian contribution to $\Bvec$.  It gives four‑point interactions, which in a statistical description contribute to the fourth‑order correlation functions of the velocity field.  Such terms are essential for producing non‑Gaussian statistics and small‑scale intermittency.\\

Altogether, the Yang–Mills Lagrangian can be summarized as:

\begin{equation}
\mathcal{L}_{\mathrm{YM}} = \underbrace{\frac12(\partial_0\Wvec)^2 - 2u^2 - 2\,\uvec\cdot(\rvec\cdot\nabla\uvec) - \frac12(\rvec\cdot\nabla\uvec)^2}_{\text{energy, helicity, enstrophy}} \;+\; \underbrace{\mathcal{L}_3 + \mathcal{L}_4}_{\text{cascade, intermittency}} 
\label{eq:quadratic-lagrangian}
\end{equation}

This explicit dictionary shows that the geometric, gauge‑invariant action naturally contains the key observables of fluid turbulence.  

\section{The Higgs Component and the Full Lagrangian}
\label{sec:ymh-lagrangian}

The Yang–Mills Lagrangian $\mathcal{L}_{\mathrm{YM}}$ describes only the tangential (rotational) degrees of freedom.  However, radial motions – local expansion and contraction of fluid parcels – also contribute to the energy cascade and intermittency.  In particular, the energy transfer between radial and tangential velocities is governed by the term $\mathcal{T}_{r\to\tau} = -\rho\,\uvec_\tau\cdot[(\uvec_r\cdot\nabla)\uvec_\tau]$ \cite{FarooqIsotropic2026}.  To incorporate these radial dynamics into a gauge‑invariant framework we introduce a Higgs sector.\\

We define a real scalar triplet $\phivec$ (the Higgs field) as the dimensionless radial velocity, scaled by the rms velocity $U_0$:

\begin{equation}
\phivec(\xvec,t) \equiv \frac{u_r(\xvec,t)}{U_0}\,\hat{\rvec}
   = \frac{\uvec(\xvec,t)\cdot\hat{\rvec}}{U_0}\,\hat{\rvec} \label{eq:phi-definition}
\end{equation}

Here $\hat{\rvec}$ is the unit vector in the radial direction from an arbitrarily chosen origin.  This choice isolates the component that breaks spherical symmetry, making $\phivec$ a natural order parameter.  In Cartesian components:

\begin{equation}
\phivec = \frac{u_r}{U_0}\begin{pmatrix} \hat{\rvec}\cdot\hat{\xvec} \\ \hat{\rvec}\cdot\hat{\yvec} \\ \hat{\rvec}\cdot\hat{\zvec} \end{pmatrix}
\end{equation}

The Higgs sector is taken to be the standard $SO(3)$ adjoint scalar theory:

\begin{equation}
\mathcal{L}_{\mathrm{Higgs}} = \frac12 |D_\mu\phivec|^2 - V(\phivec)
\qquad
V(\phivec) = \frac{\lambda}{4}\bigl(|\phivec|^2 - v^2\bigr)^2
\label{eq:Higgs-Lagrangian}
\end{equation}

where the covariant derivative is:

\begin{equation}
(D_\mu\phi)^a = \partial_\mu\phi^a + g\epsilon^{abc}W_\mu^b\phi^c
\end{equation}

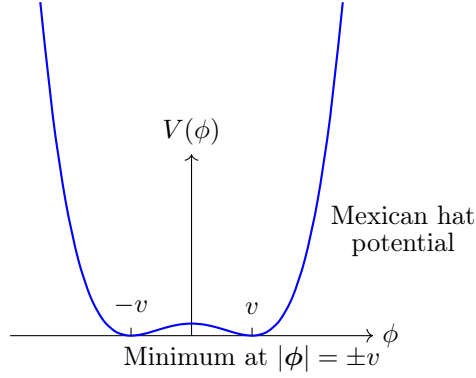
\begin{figure}[htbp]
\centering
\begin{tikzpicture}[scale=0.8]
\draw[->] (-3,0) -- (3,0) node[right] {$\phi$};
\draw[->] (0,0) -- (0,3) node[above] {$V(\phi)$};
\draw[thick, blue, domain=-2.5:2.5, smooth] plot (\x, {0.2*(\x*\x - 1)*(\x*\x - 1)});
\draw[dashed] (-1,0) -- (-1,0.2) node[above] {$-v$};
\draw[dashed] (1,0) -- (1,0.2) node[above] {$v$};
\node at (3.5,2) {Mexican hat};
\node at (3.5,1.5) {potential};
\node[below] at (1,0) {Minimum at $|\phivec| = \pm v$};
\end{tikzpicture}
\caption{The Higgs potential $V(\phivec) = \frac{\lambda}{4}(|\phivec|^2 - v^2)^2$, the ``Mexican hat'' potential.  The minima lie on the circle $|\phivec| = v$, corresponding to the vacuum expectation value (VEV) of the radial strain field $\langle u_r\rangle = v$. }
\label{fig:mexican-hat}
\end{figure}

The potential $V(\phivec)$ has a degenerate minimum at $|\phivec| = v$ (see Fig.~\ref{fig:mexican-hat}).  Choosing the vacuum expectation value (VEV) to point in the radial direction, we set in a local frame $\hat{\rvec}=\hat{\zvec}$:

\begin{equation}
\langle \phivec \rangle = v\,\hat{\zvec} = \begin{pmatrix}0\\0\\v\end{pmatrix}
\end{equation}

This VEV is invariant under rotations about the $\hat{\zvec}$-axis ($SO(2)$) but breaks the full $SO(3)$ symmetry.  The radial direction thus provides a natural symmetry‑breaking axis, with the unbroken $SO(2)$ corresponding to rotations around $\hat{\rvec}$.\\

Combining the Yang–Mills and Higgs parts we obtain the complete Lagrangian of our gauge theory:

\begin{equation}
\mathcal{L} = -\frac14 F_{\mu\nu}^a F^{a\mu\nu} + \frac12 (D_\mu\phivec)^a (D^\mu\phivec)^a - \frac{\lambda}{4}\bigl(|\phivec|^2 - v^2\bigr)^2
\label{eq:total-lagrangian}
\end{equation}

This is precisely the Lagrangian of the Georgi–Glashow model (often written for $SU(2)$, but here for $SO(3)$) – a well‑known theory that supports ’t Hooft–Polyakov monopoles \cite{Peskin-1995}.  In the next section we analyze the spontaneous symmetry breaking and show how it generates a topological mass gap, leading to the confinement of vorticity into stable, filamentary defects.\\

The translational, Galilean and $SO(3)$ gauge invariance of the Lagrangian given by Eq.~\ref{eq:total-lagrangian} is established in Appendix D.\\

In a turbulent flow the symmetry breaking occurs locally throughout the fluid.  The physical vacuum is not a single monopole but a dilute gas of defects (a monopole condensate).  When integrated over all possible origins, the net radial flux vanishes, preserving global incompressibility, while the broken $SO(2)/U(1)$ symmetry governs the local radial‑to‑tangential energy transfer – the essence of the cascade.

\subsection{Symmetry Breaking and Higgs Mechanism}
\label{sec:higgs-mechanism}

We now expand the Higgs field around its vacuum expectation value.  In the unitary gauge we rotate the Higgs fluctuation to point entirely in the third direction:

\begin{equation}
\phivec(x) \;\to\; \boldsymbol{\Sigma}(x) = 
\begin{pmatrix}
0 \\ 0 \\ v + \sigma(x)
\end{pmatrix}
\end{equation}

where $\sigma(x)$ is the physical Higgs boson field.  The two Goldstone modes have been absorbed (“eaten”) by the gauge fields.  Substituting this into the covariant derivative gives:

\begin{equation}
D_\mu \boldsymbol{\Sigma} = 
\begin{pmatrix}
g W_\mu^2\,(v+\sigma) \\[2pt]
-g W_\mu^1\,(v+\sigma) \\[2pt]
\partial_\mu \sigma
\end{pmatrix}
\end{equation}

Hence the kinetic term becomes:

\begin{align}
\frac12 |D_\mu \boldsymbol{\Sigma}|^2 
&= \frac12 (\partial_\mu \sigma)^2 
   + \frac{g^2}{2}(v+\sigma)^2\bigl[(W_\mu^1)^2+(W_\mu^2)^2\bigr] \nonumber\\
&= \frac12 (\partial_\mu \sigma)^2 
   + \frac{g^2 v^2}{2}\bigl[(W_\mu^1)^2+(W_\mu^2)^2\bigr]
   + g^2 v\sigma\bigl[(W_\mu^1)^2+(W_\mu^2)^2\bigr]
   + \frac{g^2}{2}\sigma^2\bigl[(W_\mu^1)^2+(W_\mu^2)^2\bigr]
\end{align}

The second term is a mass term for the gauge fields $W_\mu^1$ and $W_\mu^2$, while $W_\mu^3$ remains massless.  The gauge boson mass is therefore:

\begin{equation}
M_W = g v
\end{equation}

Expanding the Higgs potential around the VEV,

\begin{equation}
V(\boldsymbol{\Sigma}) = \frac{\lambda}{4}\bigl(|\boldsymbol{\Sigma}|^2 - v^2\bigr)^2
= \lambda v^2 \sigma^2 + \lambda v \sigma^3 + \frac{\lambda}{4}\sigma^4
\end{equation}

yields the Higgs mass:

\begin{equation}
M_H = \sqrt{2\lambda}\, v
\end{equation}

Since $\Wvec_\mu$ and $\phivec$ are dimensionless, the Lagrangian density has dimensions of inverse area.  The relevant parameters and their physical interpretations are collected in Table~\ref{tab:params}.

\begin{table}[htpb]
\centering
\caption{Parameters of the gauge theory and their fluid interpretation.}
\label{tab:params}
\begin{tabular}{l l l}
\toprule
\textbf{Parameter} & \textbf{Physical meaning} & \textbf{Dimension} \\
\midrule
$\eta$ & Topological microscale (core size) & $L$ \\
$v$ & Higgs VEV ($\langle u_r\rangle/U_0$) & dimensionless \\
$g = \eta^{-1}$ & Gauge coupling & $L^{-1}$ \\
$\lambda$ & Higgs self‑coupling & $L^{-2}$ \\
$M_W = g v$ & Mass of $W^{1,2}$ (inverse core radius) & $L^{-1}$ \\
$M_H = \sqrt{2\lambda}\,v$ & Mass of Higgs scalar $\sigma$ (inverse healing length) & $L^{-1}$ \\
\bottomrule
\end{tabular}
\end{table}

The original field content (three gauge fields, each with two polarizations, and three real scalars) totals nine degrees of freedom.  After symmetry breaking, the massive $W^{1,2}$ acquire a longitudinal mode (three polarisations each, total six), the massless $W^3$ retains two polarisations, and the physical Higgs $\sigma$ gives one scalar – again nine degrees of freedom (see Table~\ref{tab:dof_counting}).  This counting is preserved, as required by the Higgs mechanism.\\

\begin{table}[htpb]
\centering
\caption{Degrees of freedom before and after $SO(3)\to U(1)$ breaking.}
\label{tab:dof_counting}
\begin{tabular}{@{}lllc@{}}
\toprule
\textbf{Phase} & \textbf{Field content} & \textbf{Multiplicity $\times$ spin states} & \textbf{DoF} \\
\midrule
Before SSB & $SO(3)$ gauge fields $W^a_\mu$ & $3 \times 2$ (massless) & 6 \\
& Adjoint scalar triplet $\phivec^a$ & $3 \times 1$ & 3 \\
\midrule
After SSB & Massive $W^{1,2}_\mu$ & $2 \times 3$ (massive) & 6 \\
& Massless $W^3_\mu$ & $1 \times 2$ & 2 \\
& Physical Higgs $\sigma$ & $1 \times 1$ & 1 \\
\bottomrule
\end{tabular}
\end{table}

The massive gauge bosons $W^{1,2}$ correspond to the coherent longitudinal field $\nabla\Phi_L$.  Their mass $M_W$ implies an exponential screening length $\xi_W = M_W^{-1}$, which physically is the radius of the vortex cores.  The massless $W^3$ corresponds to the solenoidal background $\LA$; it mediates long‑range interactions and sustains the Kolmogorov cascade.  The Higgs scalar $\sigma$ represents fluctuations of the radial velocity magnitude; its correlation length $\xi_H = M_H^{-1}$ governs the healing of the radial strain outside the core.

\subsection{Geometric origin of the mass split}

The splitting of the gauge boson masses is a purely geometric effect.  When the Higgs VEV points along the $z$‑axis, $\langle\phivec\rangle = v\,\hat{\zvec}$, the $SO(3)$ symmetry is reduced to the $U(1)$ of rotations around $\hat{\zvec}$.  Rotations generated by $W^3$ leave the VEV invariant; consequently $W^3$ stays massless.  Rotations generated by $W^1$ or $W^2$ tilt the VEV away from its equilibrium direction, incurring a potential energy penalty – this is the origin of the mass $M_W$ for $W^{1,2}$.\\

\begin{figure}[htbp]
\centering
\resizebox{0.5\columnwidth}{!}{%
\begin{tikzpicture}[>=Stealth, scale=1.0, line cap=round, line join=round]
\draw[->, gray!60, thick] (0,0,0) -- (4,0,0) node[right, black] {$x\;(W^1)$};
\draw[->, gray!60, thick] (0,0,0) -- (0,4,0) node[above, black] {$z\;(W^3)$};
\draw[->, gray!60, thick] (0,0,0) -- (0,0,4) node[below left, black] {$y\;(W^2)$};
\draw[->, line width=2.5pt, red!80!black] (0,0,0) -- (0,3,0) 
    node[midway, right] {$\langle\boldsymbol{\Phi}\rangle = v\hat{\zvec}$}
    node[above] {$\phi^3$};
\filldraw[red!80!black] (0,3,0) circle (2pt);
\draw[->, thick, blue!80!black, domain=20:340, samples=60] 
    plot ({0.6*cos(\x)}, {3.3}, {0.6*sin(\x)});
\node[blue!80!black, above] at (0,3.4,0) {\small Unbroken $U(1)$};
\node[blue!80!black, right] at (0,2.2,0) {\small $W^3$ (massless)};
\draw[->, dashed, thick, purple!80, domain=-10:80, samples=30] 
    plot (0, {3*cos(\x)}, {3*sin(\x)});
\node[purple!80, right] at (-1.5,0.5,1.5) {\small $W^1$ action};
\draw[->, dashed, thick, purple!80, domain=-10:80, samples=30] 
    plot ({3*sin(\x)}, {3*cos(\x)}, 0);
\node[purple!80, left] at (2.5,0.5,-1) {\small $W^2$ action};
\node[purple!80!black, text width=3cm, align=center] at (3.2,-0.3,2) 
    {\small $W^{1,2}$ become massive};
\end{tikzpicture}%
}

\caption{Geometric origin of the mass split.  The Higgs VEV (red arrow) breaks $SO(3)\to U(1)$.  Rotations around the $z$‑axis (blue circle) leave the VEV unchanged – $W^3$ remains massless.  Rotations around the $x$ or $y$ axes (purple dashed curves) displace the VEV, incurring a potential energy penalty – $W^{1,2}$ acquire a mass $M_W = gv$.}
\label{fig:ssb_geometry}
\end{figure}
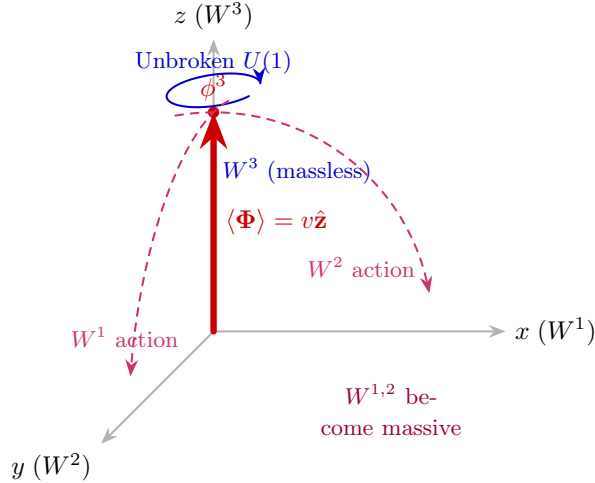

Figure~\ref{fig:ssb_geometry} illustrates this mechanism: the red arrow represents the VEV; the blue circle indicates the unbroken $U(1)$ rotations; the purple dashed curves show how $W^1$ and $W^2$ displace the VEV, leading to a mass gap.\\

This geometric locking provides a clean physical picture: the unbroken $U(1)$ sector supports the long‑range, algebraically decaying correlations of the inertial cascade, while the massive $W^{1,2}$ fields are confined to the vortex cores, giving rise to the exponential screening of enstrophy – the fluid analog of the Meissner effect.

\section{The 't Hooft-Polyakov (tHP) Monopole in the Turbulent Vacuum}
\label{sec:monopole-theory}

\subsection*{The Equations of Motion}

The total Lagrangian of the $SO(3)$ Georgi–Glashow model given by Eq.~\ref{eq:total-lagrangian} yields the following equation of motion (EOM) by varying it with respect to the gauge field $W_\mu^a$:

\begin{equation}
D_\nu F^{a\nu\mu} = g \epsilon^{abc} \phi^b D^\mu \phi^c \equiv j^{a\mu}
\label{eq:YM-eom}
\end{equation}

The right‑hand side is the Noether current associated with the global $SO(3)$ symmetry, which acts as a source for the gauge field.  Varying with respect to the Higgs field $\phi^a$ gives the covariant Klein–Gordon equation:

\begin{equation}
D_\mu D^\mu \phi^a + \lambda (|\phi|^2 - v^2) \phi^a = 0
\label{eq:Higgs-eom}
\end{equation}

Both equations are essential for the analysis of static monopole solutions.\\

In the static, time‑independent case ($\partial_0 = 0$) and in the temporal gauge $W_0^a=0$, the Gauss's law can be written as:

\begin{equation}
\partial_i E_i^a + g\,\epsilon^{abc}W_i^b E_i^c = j^{a0}
\end{equation}

In vector notation,
\begin{equation}
\nabla\cdot\Evec^a + g\,\epsilon^{abc}\,\Wvec^b\cdot\Evec^c = \rho_e^a ,\qquad \rho_e^a \equiv j^{a0}
\label{eq:vec-ampere}
\end{equation}

This is the non‑Abelian Gauss’s law.  The extra term $g\,\epsilon^{abc}\,\Wvec^b\cdot\Evec^c$ couples different colour components and is absent in electrodynamics.\\

For Ampère’s law take $\mu=i$.  The equation becomes:

\begin{equation}
\partial_\nu F^{a\nu i} + g\,\epsilon^{abc}W_\nu^b F^{c\nu i} = j^{ai}
\end{equation}

which can be written as:

\begin{equation}
 \nabla\times\Bvec^a - \partial_0\Evec^a + g\,\epsilon^{abc}\bigl(W_0^b\Evec^c + \Wvec^b\times\Bvec^c\bigr) = \jvec^a ;\qquad \jvec^a = j^{ai}
 \label{eq:amplaw}
\end{equation}

This is the non‑Abelian Ampère’s law.  The terms $g\,\epsilon^{abc}W_0^b\Evec^c$ and $g\,\epsilon^{abc}\Wvec^b\times\Bvec^c$ are the additional non‑linear interactions characteristic of a non‑Abelian gauge theory.

\subsection{The tHP Equations}

The Georgi-Glashow Lagrangian derived in the previous sections admits stable, finite‑energy topological solitons – the ’t Hooft-Polyakov (tHP) monopoles \cite{tHooft1974,Polyakov1974}.  In the context of our fluid theory, these monopoles correspond to the intense, localized vorticity filaments (“worms”) observed in high‑Reynolds‑number turbulence.  In this section we present the classical monopole solution and discuss its properties, laying the foundation for the empirical identification in Part~II.\\

We consider static configurations ($\partial_0 = 0$) in the temporal gauge $W_0^a=0$.  The energy density of the system is:

\begin{equation}
\mathcal{E} = \frac14 F_{ij}^a F_{ij}^a + \frac12 |D_i \phivec|^2 + V(\phivec)
\end{equation}

Finite energy requires that the fields approach their vacuum values at infinity:

\begin{equation}
|\phivec| \to v;\qquad D_i\phivec \to 0;\qquad F_{ij}^a \to 0\quad (r\to\infty)
\end{equation}

Moreover, the fields must be regular at the origin.  These conditions force a non‑trivial topological structure.\\

To find a spherically symmetric solution we employ the hedgehog ansatz, which locks the internal (color) space to physical space:

\begin{align}
\phi^a(\rvec) &= v\,\hat{r}^a\,H(r) \label{eq:hedgehog-phi}\\
W_i^a(\rvec) &= \epsilon^{aij}\,\hat{r}^j\,\frac{1-K(r)}{g\,r} \label{eq:hedgehog-w}
\end{align}

Here $\hat{r}^a = r^a/r$, and $H(r)$, $K(r)$ are dimensionless profile functions.  The boundary conditions for a regular, finite‑energy solution are:

\begin{equation}
H(0)=0;\quad K(0)=1;\qquad
H(\infty)=1;\quad K(\infty)=0;
\end{equation}

At the origin $H(0)=0$ locally restores the $SO(3)$ symmetry, while at infinity $H\to1$ and $K\to0$ gives the broken vacuum.\\

Substituting the ansatz into the Euler‑Lagrange equations yields a pair of coupled ordinary differential equations:

\begin{align}
r^2 H'' &= 2H K^2 + \lambda v^2 r^2 (H^2-1)H \label{eq:H-eq}\\
r^2 K'' &= K (K^2-1) + g^2 v^2 r^2 H^2 K \label{eq:K-eq}
\end{align}

These equations describe the radial structure of the monopole.\\

The system simplifies dramatically in the Bogomol’nyi–Prasad–Sommerfield (BPS) limit \cite{Bogomolnyi1976}, \cite{PrasadSommerfield1975}, where the scalar self‑coupling is set to zero while the vacuum expectation value $v$ is kept fixed\footnote{The interpretation of this limit is that $|D_\mu \boldsymbol{\phi}|^2 \;\gg\; V(\boldsymbol{\phi})$, meaning the kinetic energy of the Higgs field is far larger than the potential energy. In this limit, the scalar potential is negligible, and the energy is almost entirely contained in the gradient of the Higgs field and the gauge field.  This corresponds to a situation where the radial strain field varies slowly in space, and the self‑interaction of the strain is weak compared to its gradient energy.  In the fluid picture, this means that the energy cost of changing the radial strain from one point to another (via $D_\mu\boldsymbol{\phi}$) is much larger than the cost of having the strain away from its vacuum value.  Consequently, the defect core is dominated by the gradient energy, and the profile is determined by the balance between the gauge and scalar gradients, leading to the BPS solution.}:

\begin{equation}
\lambda \to 0,\qquad v\ \text{fixed}
\end{equation}

In this limit the potential $V(\phivec)$ vanishes, but the symmetry is still spontaneously broken by the boundary conditions.  The second‑order equations reduce to the first‑order Bogomol’nyi equations:

\begin{equation}
B_i^a = \pm D_i \phi^a
\end{equation}

where $B_i^a = \frac12 \epsilon_{ijk}F_{jk}^a$ is the non‑Abelian magnetic field.  For the hedgehog ansatz these become:

\begin{equation}
\frac{dH}{dr} = \frac{HK}{r},\qquad \frac{dK}{dr} = g^2 v^2 r H^2 - \frac{K^2-1}{r}
\end{equation}

The unique solution satisfying the boundary conditions is:

\begin{equation}
H(r) = \coth(g v r) - \frac{1}{g v r}; \qquad 
K(r) = \frac{g v r}{\sinh(g v r)} 
\end{equation}

These are the BPS monopole profiles. The BPS solution has several important features: The Higgs field $H(r)$ vanishes linearly at the origin, then rises monotonically to $1$.  The radial velocity $\uvec_r$ – which is proportional to $\phivec$ – is therefore zero at the core center, locally restoring the $SO(3)$ symmetry. The gauge field $K(r)$ decays exponentially for $r \gg (gv)^{-1}$; consequently the magnetic field (enstrophy) is confined to a region of size $\sim (gv)^{-1} = M_W^{-1}$.  This exponential screening is the non‑Abelian Meissner effect. At large distances the magnetic field assumes a Coulomb tail:

\begin{equation}
B_i^a \;\sim\; \hat{r}^a \frac{\hat{r}_i}{g r^2}
\end{equation}

which corresponds to the long‑range $U(1)$ field.  The mass of the monopole is given by the topological lower bound:

\begin{equation}
M_{\text{mon}} = \frac{4\pi v}{g}
\end{equation}

which is saturated in the BPS limit.  \\

The analytic profiles $H(r)$ and $K(r)$ derived above are precisely the functions that will be extracted from DNS data in Part~II.  We will show that the radial Higgs field $\langle u_R \rangle$ measured around isolated enstrophy maxima follows the BPS form $H(R/\eta)$ with $\eta = (gv)^{-1}$.  This provides a direct identification of the observed vortex filaments as macroscopic ’t Hooft‑Polyakov monopoles.\\

Before turning to the numerical evidence, we complete the theoretical framework by first extending the spherical tHP monoploe to strings and also deriving the Hamiltonian and the real‑space propagators that govern the energy cascade and the screening of the massive sector.  These will be the subject of the next two sections.

\section{The String Defect with 't Hooft-Polyakov Structure}
\label{sec:string-defect}

In Part II, we will empirically identify the well known vortex filaments, the "worms" of Isotropic turbulence as 't Hooft-Polyakov monopoles (Sec.~\ref{sec:monopole-validation}).  This will require that the turbulent vacuum supports not only point defects but also extended string‑like configurations.  In this section we show that the static field equations admit a cylindrically symmetric solution whose transverse cross‑section is exactly the BPS monopole profile.  This solution provides a theoretical foundation for the observed filaments and links them to the Abrikosov vortices of type‑II superconductors.\\

Consider a straight vortex filament aligned along the $z$-axis.  We work in cylindrical coordinates $(R,\varphi,z)$ with $R = \sqrt{x^2+y^2}$ and $\varphi$ the azimuthal angle.  The configuration is assumed to be:

\begin{itemize}
  \item Static: $\partial_0 = 0$,
  \item Translationally invariant along the filament: $\partial_z = 0$,
  \item Axisymmetric: $\partial_\varphi = 0$ after a suitable gauge choice.
\end{itemize}

We work in the temporal gauge $W_0^a = 0$.  The fields are taken to be:

\begin{align}
\phi^a(R,\varphi) &= v\,\hat{R}^a\,H(R) \label{eq:string-phi}\\
W_R^a &= 0, \quad W_z^a = 0 \label{eq:string-Wcomp}\\
W_\varphi^a &= \epsilon^{a\hat{z}\hat{R}}\,\frac{1-K(R)}{gR} \label{eq:string-Wvarphi}
\end{align}

Here $\hat{R}^a = (\cos\varphi,\sin\varphi,0)$ (for $a=1,2$) and $\phi^3=0$.  The antisymmetric symbol $\epsilon^{a\hat{z}\hat{R}}$ selects the color direction orthogonal to both the $z$-axis and the radial direction (it can be thought of as $\epsilon^{aij}$ with $i,j$ set to $i=\hat{R}, j=\hat{z}$). Explicitly, it gives $W_\varphi^1 = -\frac{1-K}{gR}\sin\varphi$, $W_\varphi^2 = \frac{1-K}{gR}\cos\varphi$, $W_\varphi^3=0$.  This ansatz is the direct cylindrical analog of the spherical hedgehog.\\

The static field equations for the Georgi-Glashow model are:

\begin{align}
D_i F^{aij} &= g\,\epsilon^{abc}\phi^b D_j\phi^c \label{eq:static-YM}\\
D_i D_i \phi^a &= \lambda (v^2 - |\phi|^2)\phi^a \label{eq:static-Higgs}
\end{align}

Substituting the cylindrical ansatz and using $D_i\phi^a = \partial_i\phi^a + g\,\epsilon^{abc}W_i^b\phi^c$, a  calculation similar to that of the standard tHP monopole of the last section reduces these equations to a pair of ordinary differential equations for $H(R)$ and $K(R)$:

\begin{align}
R^2 H'' &= 2H K^2 + \lambda v^2 R^2 (H^2-1)H \label{eq:Hcyl}\\
R^2 K'' &= K(K^2-1) + g^2 v^2 R^2 H^2 K \label{eq:Kcyl}
\end{align}

These are identical to the equations for the spherical monopole, with $R$ replacing the radial coordinate $r$.\\

Regularity at the origin $R=0$ requires the fields to be non‑singular.  From the ansatz, we must have:

\begin{equation}
H(0) = 0;\qquad K(0) = 1
\end{equation}

so that the Higgs field vanishes at the core (locally restoring $SO(3)$) and the gauge field is regular ($W_\varphi$ finite).  At infinity the vacuum must be recovered:

\begin{equation}
H(\infty) = 1;\qquad K(\infty) = 0
\end{equation}

giving $\phi^a \to v\hat{R}^a$ and $W_\varphi^a \to 0$.\\

In the Bogomol'nyi–Prasad–Sommerfield (BPS) limit $\lambda = g^2/2$ (i.e., $M_H=M_W$), the second‑order equations decouple into first‑order Bogomol'nyi equations:

\begin{equation}
\frac{dH}{dR} = \frac{HK}{R}\qquad 
\frac{dK}{dR} = g^2 v^2 R H^2 - \frac{K^2-1}{R}
\end{equation}

The unique solution satisfying the boundary conditions is:

\begin{align}
H(R) &= \coth(gvR) - \frac{1}{gvR} \label{eq:H-profile} \\ 
K(R) &= \frac{gvR}{\sinh(gvR)}  
\end{align}

Thus the transverse profile of the string defect is exactly the same as the spherical monopole profile, with $R$ playing the role of the radial distance.

\section{The Hamiltonian and the Helmholtz Decoupling}
\label{sec:hamiltonian}

We now derive the static energy functional of the gauge–Higgs system.  Starting from the Lagrangian $\mathcal{L}$ in Eq.~\ref{eq:total-lagrangian}, the Hamiltonian density for a static configuration ($\partial_0 = 0$) is obtained by a Legendre transformation, which for the Yang–Mills–Higgs system reduces to:

\begin{equation}
\mathcal{H}_{\text{static}} = \frac14 F_{ij}^a F_{ij}^a + \frac12 |D_i \phivec|^2 + V(\phivec)
\end{equation}

In the broken phase and at distances $r \gg \eta$ (i.e., outside the monopole cores), the fields approach their vacuum values: $\phivec \to v \hat{\rvec}$ and $V(\phivec) \to 0$.  Moreover, the gauge field becomes pure $U(1)$ in this region, $W^3_\mu$ massless, while the massive components $W^{1,2}_\mu$ are exponentially suppressed.  The Hamiltonian then splits into two distinct contributions.\\

The kinetic term for the Higgs field contains a mass‑like part when $\phivec$ is expanded around its VEV.  Writing $\phivec = v\hat{\rvec} + \text{fluctuations}$, the dominant term at long distances comes from the covariant derivative:

\begin{equation}
\frac12 |D_i \phivec|^2 \approx \frac12 g^2 v^2 |\Wvec_i \times \hat{\rvec}|^2
\end{equation}

Since $\Wvec_i$ is orthogonal to $\hat{\rvec}$ (it is built from $\Lvec = \rvec\times\uvec$), the cross product simplifies to $|\Wvec_i|^2$.  Hence we have:

\begin{equation}
\mathcal{H}_{\text{mass}} = \frac12 M_W^2 |\Wvec|^2, \qquad M_W = gv
\end{equation}

We now insert the Helmholtz–Hodge decomposition of the gauge field:

\begin{equation}
\Wvec = -\nabla \Phi_L + \nabla \times \AL
\end{equation}

into the mass term.  Since the cross term $\nabla\Phi_L \cdot (\nabla\times\AL)$ is zero (a property of Helmholtz expansion), assuming fields vanish at infinity), we obtain:

\begin{equation}
\int \mathcal{H}_{\text{mass}} \,d^3x = \frac12 M_W^2 \int \bigl( |\nabla\Phi_L|^2 + |\nabla\times\AL|^2 \bigr) d^3x
\end{equation}

The contribution from the Yang–Mills magnetic term $\frac14 F_{ij}^a F_{ij}^a$ similarly decouples into a part coming from the massless $U(1)$ sector and a part from the massive sector, but the dominant long‑range part is already accounted for by the $U(1)$ field.

\subsection{Energy partition}

The scalar potential $\Phi_L$ represents one degree of freedom (longitudinal), while the vector potential $\AL$ represents two transverse degrees of freedom (solenoidal).  In a statistically isotropic and homogeneous turbulent state, the energy stored in each degree of freedom should be equal on average (equipartition).  Consequently,

\begin{equation}
\frac{\langle |\nabla\Phi_L|^2 \rangle}{\langle |\nabla\times\AL|^2 \rangle} = \frac{1}{2}
\end{equation}

See \cite{FarooqIsotropic2026} for a detailed derivation and empirical validation.\\

This ratio translates directly into a $1:2$ partition of the kinetic energy between the coherent velocity field $\uvec_\Phi$ (derived from $\Phi_L$) and the background field $\uvec_A$ (derived from $\AL$), as will be verified against DNS data in Part~II.\\

Thus, the topological mass gap $M_W$ not only provides exponential screening of the massive sector but also enforces a strict geometric equipartition – a direct consequence of the Higgs mechanism in the broken phase.

\section{Dictionary: From Gauge Theory to Fluid Dynamics}
\label{sec:dictionary}

We now systematically translate the $SO(3)$ Georgi–Glashow Lagrangian and its equations of motion into the language of fluid mechanics.  All fields are coarse‑grained over a scale $\ell \gg \eta_K$, and we work in the spontaneously broken phase where $\langle \boldsymbol{\phi} \rangle = v\hat{\rvec}$, giving masses $M_W = gv$ to $W^{1,2}_\mu$ while $W^3_\mu$ remains massless.  

\subsection{Kinematic dictionary}
\label{sec:kinematic-dict}

We collect the fundamental identifications between gauge fields and fluid variables.  All fields are coarse‑grained over a scale $\ell \gg \eta_K$ (Kolmogorov length).  The gauge group is $SO(3)$; the index $a=1,2,3$ is a colour index.\\

The spatial components of the gauge connection are identified with the scaled specific angular momentum:

\begin{equation}
W_i^a = g^2\, (\Lvec_i)^a = g^2\, (\rvec\times\uvec)_i^a, \qquad 
g = \eta^{-1}
\end{equation}

where $\eta$ is the topological microscale (the core radius of vortex filaments) and $\rvec$ is a fixed separation vector (a parameter of the coarse‑graining).  The corresponding velocity fields reconstructed from the angular momentum decomposition (Eq.~\ref{eq:uPhi}-\ref{eq:uA},\ref{eq:totalvelocity}) are:

\begin{equation}
\uvec_\Phi = \frac{1}{r^2}(\nabla\Phi_L)\times\rvec,\qquad 
\uvec_A = \frac{1}{r^2}(\nabla\times\AL)\times\rvec,\qquad 
\uvec_r = \uvec - \uvec_\Phi - \uvec_A
\end{equation}

with $\Lvec = -\nabla\Phi_L + \nabla\times\AL$, (Eq.~\ref{eq:Ldecomp}).\\

The temporal component $W_0^a$ is not determined by the angular momentum.  In the unbroken phase we often work in the temporal gauge $W_0^a=0$.  For the broken $U(1)$ sector, consistency with the fluid equations forces:

\begin{equation}
W_0^3 = -H, \qquad H = \frac{p}{\rho} + \frac12 u^2
\end{equation}

where $H$ is the Bernoulli head.  The massive components $W_0^{1,2}$ are not needed in the bulk.\\

The real scalar triplet $\boldsymbol{\phi}$ is identified with the dimensionless radial velocity (Eq.~\ref{eq:phi-definition}):

\begin{equation}
\phi^a = \frac{u_r^a}{U_0},\qquad u_r = \uvec\cdot\hat{\rvec}
\end{equation}

Here $U_0$ is the rms velocity, and $\hat{\rvec}$ is the unit vector in the radial direction from the (arbitrary) origin of the separation vector.  In the broken phase the Higgs field acquires a vacuum expectation value:

\begin{equation}
\langle \boldsymbol{\phi} \rangle = v\,\hat{\zvec}
\end{equation}

breaking $SO(3)$ to $U(1)$.  The massive fields $W^{1,2}_\mu$ and $\phi^{1,2}$ are confined to vortex cores; outside the cores only the $U(1)$ sector ($a=3$) survives.\\

For quick reference, the identifications are:

\begin{table}
\centering
\caption{Gauge object mapping to fluid observables}
\label{tab:tt1}
\begin{tabular}{l l}
\toprule
\textbf{Gauge object} & \textbf{Fluid observable} \\
\midrule
$W_i^a$ & $g^2 (\rvec\times\uvec)^a$ \\
$W_0^3$ & $-H$ (Bernoulli head) \\
$\phi^a$ & $\uvec_r / U_0$ \\
$\langle\phi^3\rangle = v$ & $|\langle\uvec_r\rangle|/U_0$ \\
Massive $W^{1,2}_\mu$ & Confined enstrophy ($\nabla\Phi_L$) \\
Massless $W^3_\mu$ & Solenoidal background ($\nabla\times\AL$) \\
\bottomrule
\end{tabular}
\end{table}

\subsection{Reduction of the field strength to fluid observables}
\label{sec:field-strength-reduction}

Using the kinematic dictionary (Table~\ref{tab:tt1}), we now express the colour‑electric and colour‑magnetic fields in terms of fluid variables.  We work in the temporal gauge $W_0^a=0$ for simplicity; the $U(1)$ sector will later be considered with $W_0^3 = -H$. From Eq.~\ref{eq:color-magnetic} and the identification $W_k^a = g^2(\rvec\times\uvec)_k^a$, the linear part becomes:

\begin{equation}
\nabla\times\Wvec^a = g^2\nabla\times(\rvec\times\uvec) = g^2\bigl(-2\uvec - (\rvec\cdot\nabla)\uvec\bigr)
\end{equation}

The non‑Abelian part $(\frac{g}{2})\,\Wvec^b\times\Wvec^c$ is quadratic in $\uvec$ and corresponds to vortex stretching.\\

In the broken phase, far from vortex cores, the massive components $a=1,2$ are negligible.  For the massless $U(1)$ sector ($a=3$), the non‑Abelian term vanishes because $\epsilon^{3bc}$ is non‑zero only when $b,c\in\{1,2\}$ and those fields are zero in the bulk.  Moreover, the projection onto the solenoidal part of $\Wvec^3$ (i.e., the $\nabla\times\AL$ component) yields a relation (See Appendix F):

\begin{equation}
\nabla\times\Wvec^3 = \omvec
\end{equation}

where $\omvec = \nabla\times\uvec$ is the fluid vorticity.  This identification is consistent because the gradient part $-\nabla\Phi_L$ does not contribute to the curl.  Hence,

\begin{equation}
\Bvec^3 = \omvec 
\end{equation}

The electric field is defined as $E_i^a = F_{0i}^a = \partial_0 W_i^a - \partial_i W_0^a + g\,\epsilon^{abc}W_0^b W_i^c$.  In the temporal gauge ($W_0^a=0$) we would have $\Evec^a = \partial_0\Wvec^a$.  However, to recover the pressure gradient we must restore the temporal component for the $U(1)$ sector.  Using $W_0^3 = -H$ and discarding the massive $W_0^{1,2}$, we obtain:

\begin{equation}
\Evec^3 = -\nabla H - \partial_0\Wvec^3
\end{equation}

where the non‑Abelian term $g\,\epsilon^{3bc}W_0^b\Wvec^c$ vanishes because $W_0^{1,2}=0$ in the bulk.  Substituting $\Wvec^3 = g^2(\rvec\times\uvec)$ (up to a gradient that does not affect the curl) gives:

\begin{equation}
\Evec^3 = -\nabla H - \rvec\times\partial_t\uvec 
\end{equation}

Thus the colour‑electric field combines the pressure gradient and the unsteady acceleration.\\

In the broken $U(1)$ phase, the field strengths reduce to simple fluid expressions:

\begin{align}
\Bvec^3 &= \omvec, \\
\Evec^3 &= -\nabla H - \rvec\times\partial_t\uvec
\end{align}

These identifications are central to deriving the fluid equations from the Yang–Mills laws.

\subsection{Reduction to fluid equations in the broken $U(1)$ phase}

We now specialise to the broken phase where $\langle \boldsymbol{\phi} \rangle = v\hat{\zvec}$ (choosing the radial direction as the $z$‑axis).  The gauge group $SO(3)$ is spontaneously broken to $U(1)$, and the massive fields $W^{1,2}_\mu$ and $\phi^{1,2}$ are confined to the vortex cores.  Far from any defect (i.e., in the bulk of the turbulent cascade), only the massless $U(1)$ sector survives:

\[
W_\mu^3 \neq 0; \qquad W_\mu^{1,2} = 0;\qquad 
\phi^3 = v + \sigma; \qquad \phi^{1,2}=0;
\]

The Higgs field $\sigma$ represents radial fluctuations, which we will treat as a background condensate (in the leading approximation $\sigma=0$).  The coupling constants are $g = \eta^{-1}$ and $\lambda$; in the BPS limit $\lambda = g^2/2$.\\

\textbf{Gauss's law for the $U(1)$ sector}\\

For $a=3$, the non‑Abelian Gauss’s law derived in the previous subsection (Eq.~\ref{eq:vec-ampere}) becomes:

\[
\nabla\cdot\Evec^3 + g\,\epsilon^{3bc}\,\Wvec^b\cdot\Evec^c = \rho_e^3
\]

Since $b,c$ can only be $1,2$ for $\epsilon^{3bc}\neq 0$, and those fields are zero in the bulk, the second term vanishes.  Hence we have:

\[
\nabla\cdot\Evec^3 = \rho_e^3
\]

The charge density $\rho_e^3$ comes from the Higgs source:

\[
\rho_e^3 = j^{30} = g\,\epsilon^{3bc}\phi^b D^0\phi^c
\]

Again $b,c$ are restricted to $\{1,2\}$, and since $\phi^{1,2}=0$ in the bulk, one might think $\rho_e^3=0$.  However, there is a subtle contribution from the fluctuations: even though the expectation value $\langle\phi^{1,2}\rangle =0$, the covariant derivative $D^0\phi^c$ contains the gauge field $W_0^c$ which can be non‑zero.  Appendix E shows that:

\begin{equation}
\rho_e^3 = \nabla\cdot(\uvec\times\omvec) + \partial_t(\rvec\cdot\omvec)
\label{eq:gauss-current}
\end{equation}

The first term is the divergence of the Lamb vector, the second is the time derivative of the radial vorticity component.  Thus, far from cores, Gauss’s law becomes a statement about the incompressibility of the flow (pressure Poisson equation).\\

J.Z. Wu \cite{WuLambDivergence} identified the divergence of the Lamb vector \(\lvec = \uvec\times\omvec\) as a \emph{hydrodynamic charge density}, showing that it acts as the source of the Bernoulli function \(H = p/\rho + \frac12 u^2\) via the Poisson equation $\nabla^2 H = -\nabla\cdot(\uvec\times\omvec)$, (Eq~\ref{eq:bernoulli}).\\

In our gauge theory, the charge density for the massless \(U(1)\) sector given by Eq.~\ref{eq:gauss-current} simplifies for
steady flows (\(\partial_t = 0\)), and reduces exactly to Wu's hydrodynamic charge density $\rho_e = \nabla\cdot(\uvec\times\omvec)$. Thus Wu's phenomenological observation is recovered as a special case of the gauge theory, providing a direct link between the abstract \(SO(3)\) gauge structure and a well-established diagnostic of fluid dynamics.\\

\textbf{Ampère's law for the $U(1)$ sector}\\

For $a=3$, the non‑Abelian Ampère’s law (Eq.~\ref{eq:amplaw}) simplifies similarly:

\[
\nabla\times\Bvec^3 - \partial_0\Evec^3 + g\,\epsilon^{3bc}\bigl(W_0^b\Evec^c + \Wvec^b\times\Bvec^c\bigr) = \jvec^3
\]

The terms with $b,c=1,2$ involve the massive fields and are exponentially suppressed; they can be neglected in the bulk.  Hence we obtain the classical Maxwell‑like equation

\[
\nabla\times\Bvec^3 - \partial_0\Evec^3 = \jvec^3
\]

The current $\jvec^3$ is derived from the Higgs source (Eq.~\ref{eq:rho-e1}, Appendix E)):

\[
\jvec^3 = g\,\epsilon^{3bc}\phi^b \Dvec\phi^c
\]

Again, only $b,c=1,2$ contribute.  A calculation similar to the charge density yields (Eq.~\ref{eq:rho-e1}):

\[
\jvec^3 = -3\omvec - (\rvec\cdot\nabla)\omvec
\]

up to terms that vanish in the bulk.

\begin{table}[htbp]
\centering
\caption{Gauge-theoretic objects and their fluid interpretations in the vortex filament (worm) geometry.}
\label{tab:gauge-objects-worm}
\begin{tabular}{l l l l}
\toprule
\textbf{Symbol} & \textbf{Gauge Object} & \textbf{Orientation} & \textbf{Fluid Interpretation} \\
\midrule
$\Lvec_A$ & Specific angular momentum & Axial & Angular momentum density \\
$\Wvec^3$ & Massless gauge field & Axial & Unbroken $U(1)$ sector (thermal bath) \\
$\Bvec^3$ & Magnetic field & Axial & Vorticity $\omvec$ \\
$\AL$ & Vector potential & Azimuthal & Solenoidal potential (background) \\
$\Evec^3$ & Electric field & Radial & Pressure gradient + unsteady term \\
$\boldsymbol{\phi}$ & Higgs field & Radial & Radial velocity $\uvec_r$ \\
$\uvec\times\omvec$ & Lamb vector & Radial (inward) & Force per unit mass (vortex stretching) \\
\bottomrule
\end{tabular}
\end{table}

\subsection{Complete mapping table}

The following table summarizes all identifications derived above.  The left column lists the gauge‑theoretic object, the right column its fluid counterpart.\\

\begin{table}[htbp]
\centering
\footnotesize
\caption{Complete dictionary between $SO(3)$ gauge theory and turbulence observables in the symmetric (unbroken) and broken phases.  All fields are coarse‑grained.}
\label{tab:full-dictionary}
\begin{tabularx}{\textwidth}{>{\raggedright\arraybackslash}p{0.15\textwidth} >{\raggedright\arraybackslash}p{0.25\textwidth} >{\raggedright\arraybackslash}p{0.30\textwidth} >{\raggedright\arraybackslash}X}
\toprule
\textbf{Object} & \textbf{Unbroken phase ($v=0$)} & \textbf{Broken phase ($v>0$)} & \textbf{Comment} \\
\midrule
\multicolumn{4}{c}{\textit{Gauge connection}} \\
$W_i^a$ & $g^2(\rvec\times\uvec)^a$ & same & Scaled angular momentum. \\
$W_0$ & $0$ (temporal gauge) & $W_0^3 = -H$ for $U(1)$ & Bernoulli head emerges in broken phase. \\
$g=\eta^{-1}$ & Coupling constant & same & Inverse topological core radius. \\
\midrule
\multicolumn{4}{c}{\textit{Field strength}} \\
$\Bvec^a$ & $\nabla\times\Wvec^a + \frac{g}{2}\epsilon^{abc}\Wvec^b\times\Wvec^c$ & $\Bvec^3 = \omvec$; $\Bvec^{1,2}$ confined & Massive fields drop out in bulk. \\
$\Evec^a$ & $\partial_0\Wvec^a$ & $\Evec^3 = -\nabla H - \rvec\times\partial_t\uvec$ & Pressure gradient via $W_0^3$. \\
\midrule
\multicolumn{4}{c}{\textit{Sources}} \\
$\jvec^a$ & $g\epsilon^{abc}\phi^b \Dvec\phi^c$ & $\jvec^3 = -3\omvec - (\rvec\cdot\nabla)\omvec$ & Current $\sim$ vorticity + gradient. \\
$\rho_e^a$ & $g\epsilon^{abc}\phi^b D^0\phi^c$ & $\rho_e^3 = \nabla\cdot(\uvec\times\omvec) + \partial_t(\rvec\cdot\omvec)$ & Charge $\sim$ Lamb vector. \\
\midrule
\multicolumn{4}{c}{\textit{Symmetry breaking sector}} \\
$\boldsymbol{\phi}$ & $\uvec_r/U_0$ & $\langle\phi^3\rangle=v$; $\phi^{1,2}$ confined & VEV breaks $SO(3)\to U(1)$. \\
Massive $W^{1,2}_\mu$ & -- & Confined enstrophy ($\nabla\Phi_L$) & Yukawa decay $e^{-M_W r}$. \\
Massless $W^3_\mu$ & -- & Solenoidal background ($\nabla\times\AL$) & Long‑range $1/r^2$ tail. \\
Physical Higgs $\sigma$ & -- & Fluctuations of $|\uvec_r|$ & Mass $M_H=\sqrt{2\lambda}v$. \\
\bottomrule
\end{tabularx}
\end{table}

\paragraph{Note on dimensions.}
In the equations above, we have set $\eta = 1$.  The full \(\eta\)-dependent forms are:

\begin{align}
\Wvec &= \eta^{-2}\,\rvec\times\uvec, \\
g &= \eta^{-1}, \\
\nabla\times\Wvec^3 &= \eta^{-2}\,\boldsymbol{\omega}, \\
\Evec^3 &= -\nabla H - \eta^{-2}\,\rvec\times\partial_t\uvec.
\end{align}

The physical values of $\eta$ are given in Sec.~\ref{sec:monopole-validation}.\\

This dictionary provides a direct, quantitative link between the abstract gauge theory and the fluid quantities measured in DNS.  It is used extensively in the validation sections (Secs.~\ref{sec:massgap-validation} and~\ref{sec:monopole-validation}) to extract the mass gap and the monopole profile.

\subsection{Generalized Fluid Equation from Gauss's and Ampère's Laws}
\label{sec:generalized-fluid-equation}

In the broken phase, the Yang–Mills equations for the massless $U(1)$ sector reduce to two equations: Gauss's law and Ampère's law.  Using the kinematic dictionary from Sec.~\ref{sec:dictionary}, these equations can be expressed entirely in terms of fluid variables.  The result is a generalized fluid equation that contains the Euler equation as a special limit.\\

Ampère's law for the massless $U(1)$ sector is:

\begin{equation}
\nabla\times\mathbf{B}^3 - \partial_0\mathbf{E}^3 
+ g\,\epsilon^{3bc}\bigl(W_0^b\,\mathbf{E}^c + \mathbf{W}^b\times\mathbf{B}^c\bigr) = \mathbf{j}^3
\end{equation}

In the broken phase, the massive fields $W^{1,2}$ are exponentially suppressed outside the vortex cores.  In the bulk ($r \gg \eta$), we have $W_0^b \approx 0$ and $\mathbf{W}^b \approx 0$ for $b=1,2$.  The non-Abelian commutator term therefore vanishes, and Ampère's law reduces to:

\begin{equation}
\nabla\times\mathbf{B}^3 - \partial_0\mathbf{E}^3 = \mathbf{j}^3.
\end{equation}

Substituting the dictionary expressions:

\begin{equation}
\mathbf{B}^3 = \omvec; \qquad 
\mathbf{E}^3 = \nabla H - \eta^{-2}\,\rvec\times\partial_t\uvec; \qquad 
\mathbf{j}^3 = -3\omvec - (\rvec\cdot\nabla)\omvec;
\end{equation}

we obtain:

\begin{equation}
\nabla\times\omvec - \partial_0\left(\nabla H - \eta^{-2}\,\rvec\times\partial_t\uvec\right)
= -3\omvec - (\rvec\cdot\nabla)\omvec
\end{equation}

Expanding the time derivative:

\begin{equation}
\nabla\times\omvec - \nabla(\partial_t H) + \eta^{-2}\,\rvec\times\partial_t^2\uvec
= -3\omvec - (\rvec\cdot\nabla)\omvec
\end{equation}

Using the incompressibility condition $\nabla\cdot\uvec = 0$, we have the identity:

\begin{equation}
\nabla\times\omvec = -\nabla^2\uvec
\end{equation}

Substituting this gives:

\begin{equation}
-\nabla^2\uvec - \nabla(\partial_t H) + \eta^{-2}\,\rvec\times\partial_t^2\uvec
= -3\omvec - (\rvec\cdot\nabla)\omvec 
\label{eq:generalized-fluid}
\end{equation}

This equation is an exact consequence of Ampère's law and the kinematic dictionary within the coarse‑grained effective theory.  It is structurally similar to the Euler equation in rotational form $\partial_t\uvec + (\uvec\cdot\nabla)\uvec = -\nabla H + \uvec\times\omvec$, but contains additional terms involving the separation vector $\rvec$ and second-order time derivatives.  These extra terms encode the angular momentum dynamics that are absent in the classical Euler description.  It may be possible to recover the Euler's equation by suitable coarse graining techniques and will be the subject of future studies.\\

We note that Marmanis \cite{Marmanis1998} had also cast the equations of fluid motion into Maxwellian form.

\newpage

\part{Applications}

\section{First Validation: Connection to Wave Turbulence}
\label{sec:zakharov}

In the symmetric phase ($v=0$), all three color components of the gauge field are massless.  The Yang–Mills Lagrangian then describes a system of weakly nonlinear waves.  We show that in this limit the Hamiltonian reduces to the canonical three‑wave form studied by Zakharov, L'vov, and Falkovich~\cite{ZLF1992}, thereby linking our gauge theory to the extensive literature on wave turbulence.\\

Working in the temporal and Coulomb gauges ($W_0^a=0$, $\partial_i W_i^a=0$), the quadratic part of the Hamiltonian is:

\begin{equation}
H_0 = \frac12 \int d^3x \left[ \Pi_i^a \Pi_i^a + \frac12 (\partial_i W_j^a - \partial_j W_i^a)^2 \right]
\end{equation}

where $\Pi_i^a = \partial_0 W_i^a$ are the conjugate momenta.  Expanding in Fourier modes,

\begin{equation}
W_i^a(\xvec) = \frac{1}{\sqrt{V}} \sum_{\kvec,\lambda} \varepsilon_i^\lambda(\kvec) \left( a_\lambda^a(\kvec) e^{i\kvec\cdot\xvec} + \text{c.c.} \right)
\end{equation}

with $\varepsilon_i^\lambda$ transverse polarisation vectors, one obtains:

\begin{equation}
H_0 = \sum_{\kvec,\lambda,a} \omega_{\kvec}\, a_\lambda^{a*}(\kvec) a_\lambda^a(\kvec)
\end{equation}

where $\omega_{\kvec} = U_0 |\kvec|$ and $U_0$ is the characteristic sweeping velocity.  Thus the free theory consists of three copies of massless wave modes with a linear dispersion relation.\\

The cubic part of the Hamiltonian comes from the non‑Abelian term in the field strength:

\begin{equation}
H_3 = \frac{g}{2} \int d^3x \; \epsilon^{abc} (\partial_i W_j^a - \partial_j W_i^a) W_i^b W_j^c
\end{equation}

Substituting the Fourier expansion and integrating over space enforces wavevector conservation $\kvec + \pvec + \qvec = \mathbf{0}$.  The result can be written as:

\begin{equation}
H_3 = \frac{1}{3!} \sum_{\substack{\kvec,\pvec,\qvec \\ \kvec+\pvec+\qvec=0}} 
      V_{\kvec\pvec\qvec} \; a^*(\kvec) a(\pvec) a(\qvec) + \text{c.c.}
\end{equation}

where $a(\kvec)$ collectively denotes the colour and polarisation components.  The vertex $V_{\kvec\pvec\qvec}$ is linear in the wavevectors (because of the derivatives) and contains the antisymmetric colour factor $\epsilon^{abc}$.\\

In a statistically isotropic ensemble, the colour indices can be averaged over, leading to an effective scalar amplitude $A_{\kvec}$.  The interaction Hamiltonian then takes the canonical form:

\begin{equation}
H_3 = \frac12 \sum_{\kvec,\pvec,\qvec} \mathcal{V}_{\kvec\pvec\qvec} \,
      A_{\kvec}^* A_{\pvec} A_{\qvec} \, \delta_{\kvec,\pvec+\qvec} + \text{c.c.}
\end{equation}

with $\mathcal{V}_{\kvec\pvec\qvec}$ homogeneous of degree $1$ in the wavevectors.  This is precisely the structure required for the Zakharov conformal transformation, which leads to the Kolmogorov‑Zakharov energy spectrum $E(k) \sim k^{-5/3}$ for a direct cascade.

\subsection{Comparison with phenomenological wave turbulence}

In standard wave turbulence theory, the Hamiltonian $H = H_0 + H_3$ is postulated on phenomenological grounds, with the dispersion relation and the vertex taken from the specific physical system.  In our framework, the cubic vertex is not an ad‑hoc addition: it is an unavoidable consequence of local $SO(3)$ gauge invariance.  The Yang–Mills action forces the non‑Abelian self‑interaction to appear exactly as required for three‑wave interactions.  Thus the emergence of the Zakharov structure provides a geometric origin for wave turbulence, showing that the observed cascades are manifestations of non‑Abelian gauge symmetry.\\

This connection is a consistency check: our gauge theory reproduces a known result of weak‑turbulence theory in the symmetric phase.  In the broken phase, the mass gap modifies the dispersion relation and the interaction vertices, leading to the confinement phenomena discussed in the next three sections.

\section{Second Validation: Spectral Partitioning and the Mass Gap}
\label{sec:massgap-validation}

We now test the most direct prediction of the broken phase: the energy partition between the massive and massless sectors.  Using DNS data of forced isotropic turbulence from Li et al. \cite{Li2008} ( $Re_\lambda \approx 433$, grid cutout $N=512^3$, relevant parameters of the simulation summarized in Table~\ref{tab:extracted-parameters}), we decompose the velocity field according to the $\Lvec$ framework described in Sec.~\ref{sec:Lvec-framework}.  The fields $\Lvec_\Phi$ (coherent, massive sector) and $\Lvec_A$ (background, massless sector) are obtained via Helmholtz decomposition of the specific angular momentum $\Lvec $, using a fixed separation $|\rvec|$ within the inertial range.\\

The three‑dimensional Edicity spectra (also see Eq.~\ref{eqn:editicity-def}) are defined as:

\begin{align}
E_\Phi(k) &= \frac12 \oint_{|\kvec|=k} \langle \hat{\Lvec}_\Phi(\kvec) \cdot \hat{\Lvec}_\Phi^*(\kvec) \rangle \, dS(\kvec) \\
E_A(k) &= \frac12 \oint_{|\kvec|=k} \langle \hat{\Lvec}_A(\kvec) \cdot \hat{\Lvec}_A^*(\kvec) \rangle \, dS(\kvec)
\end{align}

\begin{figure}[htbp]
\centering
\begin{subfigure}[b]{0.55\linewidth}
    \centering
    \includegraphics[width=\linewidth]{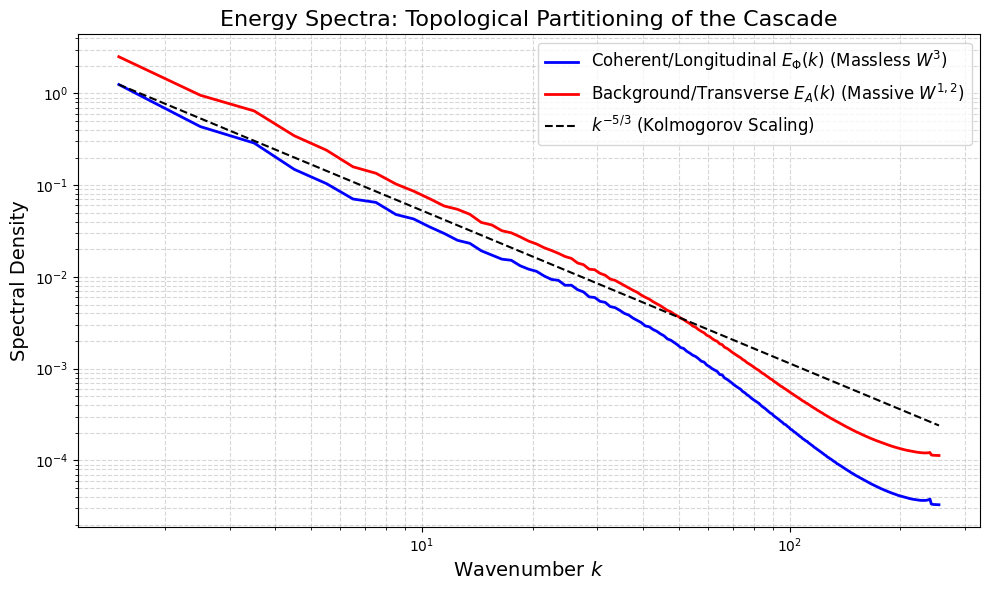}
    \caption{Spectra $E_\Phi(k)$ (blue) and $E_A(k)$ (red).  Both follow $k^{-5/3}$.}
\end{subfigure}
\hfill
\begin{subfigure}[b]{0.55\linewidth}
    \centering
    \includegraphics[width=\linewidth]{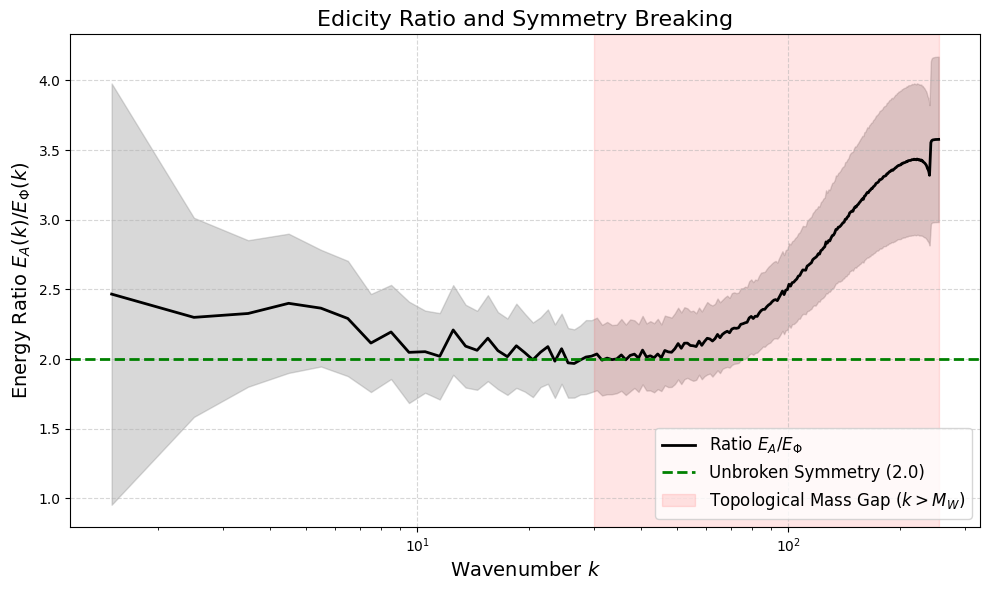}
    \caption{Ratio $E_A/E_\Phi = 2$ for $k<40$, diverging at $k \approx 40$.}
\end{subfigure}
\caption{(a) Edicity spectra showing the Kolmogorov scaling. (b) The edicity ratio remains constant at $2.0$ in the inertial range, then diverges sharply at $M_W \approx 40$, the topological mass gap.  Data from the Johns Hopkins Turbulence Database~\cite{Li2008}.}
\label{fig:massgap}
\end{figure}

where hats denote Fourier transforms.  In the inertial range both spectra follow the Kolmogorov $k^{-5/3}$ scaling (Fig.~\ref{fig:massgap}(a)).  The ratio:

\begin{equation}
\mathcal{R}(k) = \frac{E_\Phi(k)}{E_A(k)}
\end{equation}

is constant over the inertial range and equals $1/2$ (Fig.~\ref{fig:massgap}(b)).  This $1:2$ equipartition is exactly what the Hamiltonian predicts: one degree of freedom in the longitudinal sector ($\nabla\Phi_L$) and two degrees of freedom in the solenoidal sector ($\nabla\times\AL$).\\

The ratio \(\mathcal{R}(k)\) is computed for each of the 20 random origins (see \cite{FarooqIsotropic2026} for more details on numerical methodology) and then averaged.  The error bars in Fig.~\ref{fig:massgap}(b) represent the standard deviation across origins, confirming that the \(1:2\) partition is statistically robust.\\

As $k$ increases towards the dissipation range, the ratio $\mathcal{R}(k)$ remains flat until a critical wavenumber $k_*$, then rises sharply (Fig.~\ref{fig:massgap}(b)).  This divergence marks the point where the observation scale penetrates the vortex cores.  Inside the cores, the radial strain vanishes, $\langle\phivec\rangle \to 0$, and the full $SO(3)$ symmetry is locally restored.  Consequently, the massive $W^{1,2}$ bosons are no longer suppressed and the energy partition breaks down.\\

The transition occurs at:

\begin{equation}
M_W \equiv k_* \approx 40 \quad (\text{in domain units})
\end{equation}

In domain units, where the domain length is \(2\pi\), this corresponds to a physical scale:

\begin{equation}
\lambda_* = \frac{2\pi}{k^*} \approx 0.126\text{--}0.157 \quad \text{(domain units)}.
\end{equation}

This scale is well within the dissipation range, as we will confirm in the next section.\\

The observation of a strict \(1:2\) equipartition over a decade of wavenumbers, followed by a sharp break at \(M_W \approx 40\), provides strong evidence that the turbulent vacuum is in a spontaneously broken \(SO(3)\) phase.  The mass gap \(M_W\) acts as a UV cutoff that separates the scale-invariant inertial cascade from the symmetry-restored cores.  In the next section, we examine the real-space structure of these cores and identify them as 't Hooft-Polyakov monopoles.

\section{Third Validation: The Macroscopic tHP Monopole}
\label{sec:monopole-validation}

Having identified the macroscopic boundary of the vacuum in momentum space ($M_W \sim 40$), we now turn to real space.  If the geometric suppression of the inertial cascade is truly driven by the fluid Higgs mechanism, the regions where the vacuum symmetry is locally restored cannot be random fluctuations; they must be highly structured, stable non‑Abelian defects.\\

The Georgi–Glashow formalism dictates a specific geometric signature for the 't Hooft–Polyakov (tHP) monopole: at the defect core, the scalar field (the radial strain) must vanish to avoid a singularity, locally restoring the full $SO(3)$ symmetry.  Simultaneously, the non‑Abelian magnetic field (the enstrophy) reaches a maximum and is confined by the massive $W^{1,2}$ gauge fields.\\

High‑resolution DNS of isotropic turbulence have long revealed the spontaneous emergence of intense, thin vortex filaments – the ``worms'' \cite{VincentMeneguzzi1991,Jimenez1993}.  These filaments are characterized by extreme localized vorticity, diameters of the order of the Kolmogorov length, and remarkable stability despite being embedded in a chaotic background.  Historically treated as statistical outliers responsible for intermittency, they possess exactly the kinematic profile predicted by our gauge theory.  \\

The classical tHP monopole is a spherically symmetric point defect.  However, in a turbulent flow, background straining and the alignment of enstrophy along the vorticity direction extrude the defect into a quasi‑one‑dimensional string. We therefore identify these worms as the macroscopic realisation of tHP monopole threads (Sec.\ref{sec:string-defect}).  The cross‑section of this string retains the monopole topology, as illustrated in Fig.~\ref{fig:topological_vacuum}.

\begin{figure}[H]
\centering
\resizebox{0.9\textwidth}{!}{%
\begin{tikzpicture}[scale=1.0, every node/.style={scale=0.9}]

    \node at (-2.3, 6.5) {\textbf{(a) Gauge-Higgs Symmetry Breaking}};
    
    \begin{scope}[shift={(-3, 1.5)}]
        \draw[thick, ->] (-2.5, 0) -- (2.5, 0) node[right] {$\phi_1$};
        \draw[thick, ->] (0, -1.5) -- (0, 3.5) node[above] {$V(\boldsymbol{\phi})$};
        \draw[thick, ->] (-1.5, -1) -- (3, 2) node[right] {$\phi_2$};
        \draw[domain=0:360, samples=80, variable=\t, thick, fill=blue!10, opacity=0.8] 
            plot ({2*cos(\t)}, {0.5*sin(\t) + 0.8});
        \draw[domain=-1.2:1.2, samples=50, variable=\x, thick, blue] 
            plot ({\x}, {(\x*\x - 1)*(\x*\x - 1) + 0.2});
        \draw[domain=-1.5:1.5, samples=50, variable=\x, thick, blue, dashed] 
            plot ({\x}, {(\x*\x - 1)*(\x*\x - 1) + 0.2});
        \filldraw[red] (1, 0.2) circle (2pt) node[above, right] {VEV ($\langle \uvec_r \rangle \neq 0$)};
        \draw[->, thick, red] (1, 0.2) -- (1, 1.5) node[midway, right] {Massive $W^{1,2}$};
        \draw[->, thick, green!60!black] (1, 0.2) arc (0:60:1cm and 0.25cm);
        \node[green!60!black, below] at (0.5, 0) {Massless $W^3$ (Bath)};
    \end{scope}

    \node at (4.5, 6.5) {\textbf{(b) The Topological Fluid Vacuum}};

    \begin{scope}[shift={(1.5, 0)}]
        \draw[thick, fill=gray!5] (0,0) -- (4,0) -- (5,1.5) -- (1,1.5) -- cycle;
        \draw[thick, fill=gray!5, opacity=0.8] (0,0) -- (0,4) -- (4,4) -- (4,0) -- cycle;
        \draw[thick, fill=gray!10, opacity=0.8] (4,0) -- (5,1.5) -- (5,5.5) -- (4,4) -- cycle;
        \draw[thick, fill=gray!15, opacity=0.8] (0,4) -- (1,5.5) -- (5,5.5) -- (4,4) -- cycle;
        \foreach \i in {0.5, 1.5, 2.5, 3.5} {
            \draw[cyan, thick, opacity=0.4, decorate, decoration={snake, amplitude=1mm, segment length=5mm}] (\i, 0.5) -- (\i+0.5, 4.5);
            \draw[cyan, thick, opacity=0.4, decorate, decoration={snake, amplitude=1mm, segment length=5mm}] (0.5, \i) -- (4.5, \i+0.5);
        }
        \draw[red, line width=4pt, opacity=0.9, smooth, tension=0.7] plot coordinates {(1, 0.5) (1.2, 2) (1.8, 3.5) (1.5, 5)};
        \draw[orange, line width=1.5pt, smooth, tension=0.7] plot coordinates {(1, 0.5) (1.2, 2) (1.8, 3.5) (1.5, 5)};
        \draw[red, line width=3.5pt, opacity=0.9, smooth, tension=0.6] plot coordinates {(3.5, 1) (3.0, 2.5) (3.8, 4) (4.2, 5.2)};
        \draw[orange, line width=1.5pt, smooth, tension=0.6] plot coordinates {(3.5, 1) (3.0, 2.5) (3.8, 4) (4.2, 5.2)};
        \draw[red, line width=3pt, opacity=0.9, smooth cycle, tension=0.8] plot coordinates {(2, 1.5) (2.5, 2) (2.2, 2.8) (1.5, 2.2)};
        \draw[orange, line width=1pt, smooth cycle, tension=0.8] plot coordinates {(2, 1.5) (2.5, 2) (2.2, 2.8) (1.5, 2.2)};
        \node[align=center, black!80!black, font=\footnotesize] at (2.5, 0.5) {Broken $U(1)$ Sector \\ Massless Sweeping Bath \\ $\AL$ (Power-Law)};
        \draw[<-, thick] (1.8, 3.5) -- (4.5, 3.5) node[right, align=left, font=\footnotesize] {Unbroken $SO(3)$ Sector\\ Massive Defect Core ($\boldsymbol{\Phi}_L$) \\ (tHP Monopole)};
        \draw[dashed, thick] (3.0, 2.5) circle (0.3);
        \draw[dashed, thick] (3.2, 2.7) -- (5.5, 2.0);
        \begin{scope}[shift={(6.5, 1.5)}]
            \draw[fill=white, thick] (0,0) circle (1);
            \draw[fill=red!20, draw=red, thick] (0,0) circle (0.5);
            \filldraw[red] (0,0) circle (2pt);
            \node[font=\scriptsize] at (0, 0.7) {$\langle \uvec_r \rangle \neq 0$};
            \node[font=\scriptsize, align=center] at (0.4, -0.2) {Core:\\ $\langle \uvec_r \rangle \to 0$};
        \end{scope}
    \end{scope}
\end{tikzpicture}
}
\caption{\textbf{(a)} Spontaneous symmetry breaking of the $SO(3)$ gauge group via condensation of the radial strain field (Mexican hat potential).  \textbf{(b)} The resulting topological phase partition: a continuous, massless $U(1)$ bath (wavy lines) pierced by discrete, massive 't Hooft-Polyakov monopole threads (red filaments).  The inset shows the cross‑section: the radial strain $\langle \uvec_r \rangle$ vanishes at the core, locally restoring $SO(3)$.}
\label{fig:topological_vacuum}
\end{figure}

\begin{figure}[htbp]
\centering
\resizebox{0.4\textwidth}{!}{%
\begin{tikzpicture}[>=Stealth, line cap=round, line join=round]
    \colorlet{wormcolor}{blue!15}
    \colorlet{axiscolor}{black}
    \colorlet{rcolor}{gray!80}
    \colorlet{Rcolor}{red!80!black}
    \colorlet{urotcolor}{blue!80!black}
    \colorlet{urcolor}{red!80!black}
    \filldraw[fill=wormcolor, draw=blue!50, thick] (0, -3) ellipse (2 and 0.6);
    \draw[blue!50, thick, dashed] (2, 3) arc (0:180:2 and 0.6);
    \fill[wormcolor, opacity=0.7] (-2,-3) rectangle (2,3);
    \draw[blue!50, thick] (-2,-3) -- (-2,3);
    \draw[blue!50, thick] (2,-3) -- (2,3);
    \draw[blue!50, thick] (-2, 3) arc (180:360:2 and 0.6);
    \draw[->, thick, axiscolor] (0, -3.5) -- (0, 4) node[above] {$\hat{\mathbf{n}} \parallel \omvec_c$};
    \filldraw[black] (0,0) circle (2pt) node[left=2pt] {$\xvec_c$};
    \coordinate (X) at (2, 1);
    \filldraw[black] (X) circle (2pt) node[right=2pt] {$\xvec$};
    \draw[->, thick, rcolor, dashed] (0,0) -- (X) node[midway, above left=-2pt] {$\rvec$};
    \coordinate (Proj) at (0, 1);
    \draw[dashed, axiscolor!50] (0,0) -- (Proj);
    \draw[->, very thick, Rcolor] (Proj) -- (X) node[midway, below] {$\Rvec$};
    \draw[->, very thick, urotcolor] (X) ++(0,0.1) arc (0:60:2 and 0.6) node[above right] {$\uvec_{\text{rot}} = \uvec_A$ (Unbroken $U(1)$)};
    \draw[->, very thick, urcolor] (X) -- ++(1.5, 0) node[right] {$\uvec_r$ (Broken $W^{1,2}$)};
\end{tikzpicture}
}
\caption{Cylindrical coordinate system around a vortex core.  The separation vector $\rvec$ is projected onto the plane orthogonal to the vorticity axis $\hat{\mathbf{n}}$ to obtain the cylindrical radius vector $\Rvec$ (with magnitude $R$).  The velocity decomposes into a solenoidal (rotational) part $\uvec_{\text{rot}} = \uvec_A$ (massless $U(1)$ sector, the thermal bath boson) and a radial part $\uvec_r$ (massive Higgs sector, the worm boson).  The identity $\nabla\times\mathbf{W}^3 = \omvec$ (Appendix F) justifies this decomposition.}
\label{fig:worm_schematic}
\end{figure}
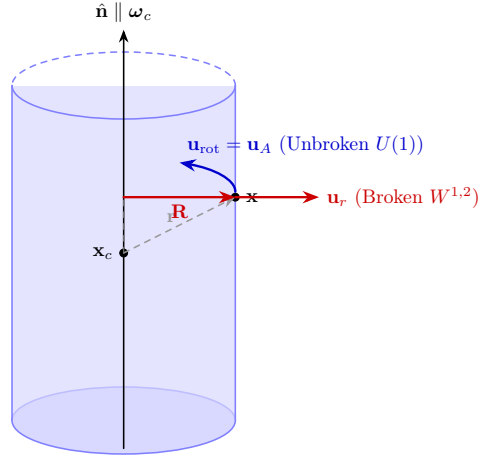

The key identification \(\nabla\times\mathbf{W}^3 = \omvec\) ensures that the solenoidal velocity \(\uvec_A\) carries the vorticity, while the remaining part \(\uvec_r\) is the radial (dilatational) component associated with the Higgs field.  This decomposition is illustrated in Fig.~\ref{fig:worm_schematic}.

\subsection{Numerical methodology for the extraction of the Higgs profile}

To validate that the ``worms" are indeed tHP threads, we need to plot and compare with the theoretical Higgs field profile (Eq.~\ref{eq:H-profile}).  Here we need to first identify the location of the tHP threads and then extract the value of the Higgs field $\phivec$ at a radial distance $R$ from the ``center" of the tHP thread and see if it conforms with Eq.~\ref{eq:H-profile}. We will perform this analysis by sampling a set of $50$ points with the most intense enstrophy within the isotropic field as the selected cores where we will compute the Higgs field.  It may be noted that Eq.~\ref{eq:H-profile} has a paramter $\eta$ which is not known a priori.  Therefore we will have to find the value of $\eta$ which gives a best fit to the theoretical profile.  \\

We use DNS data from the \cite{Li2008} ($Re_\lambda \approx 433$, grid $512^3$, forced isotropic turbulence).  The extraction procedure follows the L-field algorithm and consists of the following steps.\\

The vorticity field \(\omvec = \nabla \times \uvec\) is computed using a Fourier spectral method.  The enstrophy density \(|\omvec|^2\) is then evaluated at each grid point.  Local maxima of enstrophy are identified as candidate vortex cores (worms).  To ensure statistical independence of the extracted structures, a minimum separation of 15 grid points (approximately \(0.09\) domain units) is enforced between cores, yielding \(N = 50\) isolated cores.\\

For each core located at \(\xvec_c\), we extract a local sub-volume of radius \(R_{\max} = 20\,\Delta x\), where \(\Delta x\) is the grid spacing.  The local separation vector is:

\begin{equation}
\rvec = \xvec - \xvec_c; \qquad |\rvec| \le R_{\max}
\end{equation}

The core vorticity direction (see Fig.~\ref{fig:worm_schematic}) defines the local symmetry axis:
\begin{equation}
\hat{\mathbf{n}} = \frac{\omvec_c}{|\omvec_c|}
\end{equation}

The separation vector is projected onto the plane orthogonal to \(\hat{\mathbf{n}}\) to obtain the cylindrical radius vector:

\begin{equation}
\Rvec = \rvec - (\rvec\cdot\hat{\mathbf{n}})\hat{\mathbf{n}}, \qquad R = |\Rvec|
\end{equation}

This projection is essential because the theoretical BPS profile \(H(R/\eta)\) is defined in the plane perpendicular to the filament axis (see Sec.~\ref{sec:string-defect}).  The azimuthal angle \(\varphi\) is not needed due to the assumed axisymmetry of the monopole string.

\subsubsection*{L-field reconstruction of the solenoidal velocity}

The relative velocity in the core rest frame is:
\begin{equation}
\uvec_{\text{rel}}(\xvec) = \uvec(\xvec) - \uvec(\xvec_c)
\end{equation}

The local angular momentum (impulse) field is:

\begin{equation}
\Lvec = \Rvec \times \uvec_{\text{rel}}
\end{equation}

From the definition of the solenoidal velocity in terms of the vector potential (see Eq.~\ref{eq:AL-to-u_A}),

\begin{equation}
\uvec_A = \frac{1}{R^2}(\nabla\times\mathbf{A}_L)\times\Rvec
\end{equation}

and using \(\mathbf{W}^3 = \nabla\times\mathbf{A}_L\), (see Eq.~\ref{eq:W3-AL}), we reconstruct the rotational part of the velocity as:

\begin{equation}
\uvec_{\text{rot}} \equiv \uvec_A = \frac{\Lvec \times \Rvec}{R^2}
\end{equation}

The identity \(\nabla\times\mathbf{W}^3 = \omvec\) (Appendix F) ensures that this reconstruction correctly isolates the solenoidal (vorticity-carrying) part of the flow.  The radial (dilatational) residual is then:
\begin{equation}
\uvec_r = \uvec_{\text{rel}} - \uvec_{\text{rot}}
\end{equation}

The scalar radial component, which serves as the empirical Higgs field, is obtained by projecting onto the radial direction:
\begin{equation}
u_R(\xvec) = \uvec_r(\xvec) \cdot \hat{\Rvec}, \qquad \hat{\Rvec} = \frac{\Rvec}{R}
\end{equation}

At the core center (\(R=0\)), the geometry is singular; we enforce \(u_R(0) = 0\) to impose the symmetry restoration condition \(\langle \uvec_r \rangle \to 0\) at the origin.

\paragraph{Numerical extraction of the BPS profile.}
For each isolated vortex core located at $\mathbf{x}_c$, we extract a local sub-volume of radius $R_{\max} = 20\,\Delta x$ grid points, with $\Delta x =2\pi/1024$ give $R_{\max} \approx 0.1227$. The local grid is defined by the integer coordinates: $i,j,k \in [-20, 20]$ which is $68,\!921 \; \text{grid points}$.\\

However, to maintain a consistent spherical geometry, we apply a spherical mask $
i^2 + j^2 + k^2 \le 20^2$ which retains approximately $42,\!000$ points per core.  With $N=50$ isolated cores, the total number of data points used in the ensemble average is $N_{\text{total}} \approx 50 \times 42,\!000 = 2.1 \times 10^6 \quad \text{points}$.\\

For each point, we compute the radial component $u_R = (\mathbf{u}_{\text{rel}} - \mathbf{u}_{\text{rot}}) \cdot \hat{\mathbf{R}}$ and bin the data in cylindrical radius $R$.  The ensemble average $\langle u_R \rangle$ is then computed by averaging over all cores in each radial bin.  This large sample size ensures that the random background fluctuations cancel, leaving only the coherent topological signal.  The resulting profile is then fitted to the BPS function $H(R/\eta) = \coth(R/\eta) - \eta/R$ using non-linear least squares. \\

The data are binned into radial bins of width $\Delta R = \Delta x/2 \approx 0.00307$ domain units, ranging from $R=0$ to $R_{\max} \approx 0.1227$ domain units.  The number of bins is:

\begin{equation}
N_{\text{bins}} = \frac{R_{\max}}{\Delta R} = \frac{20\,\Delta x}{\Delta x/2} = 40
\end{equation}

The ensemble average in each bin is computed as:

\begin{equation}
\langle u_R \rangle(R_i) = \frac{1}{N_{\text{cores}}} \sum_{c=1}^{N_{\text{cores}}} \frac{1}{N_{c,i}} \sum_{p \in \text{bin } i} u_R^{(c)}(\mathbf{x}_p)
\end{equation}

where $N_{c,i}$ is the number of points from core $c$ that fall into bin $i$.

\subsubsection*{VEV extraction and normalization}

The vacuum expectation value (VEV) \(v\) is extracted from the asymptotic tail of the profile, where the topological structure merges into the homogeneous background:

\begin{equation}
v = \frac{1}{U_0}\lim_{R \to \infty} \langle |u_R| \rangle(R)
\end{equation}

where $U_0$ is the RMS velocity of the turbulent field. In practice, \(v\) is obtained by averaging the last 10 radial bins (at the largest \(R\) values).  The profile is then normalized:

\begin{equation}
\phi_{\text{emp}}(R) = \frac{\langle |u_R| \rangle(R)}{v}
\end{equation}

\subsubsection*{Theoretical comparison}

The normalized profile is fitted to the theoretical BPS function:

\begin{equation}
H\left(\frac{R}{\eta}\right) = \coth\left(\frac{R}{\eta}\right) - \frac{\eta}{R}
\end{equation}

where \(\eta \) is the topological microscale.  The fit is performed using non-linear least squares over the range \(R < R_{\max}/2\) to avoid the noisy tail.  The extracted \(\eta\) is compared to the value obtained from the spectral mass gap measurement (Sec.~\ref{sec:massgap-validation}), providing a consistency check between the real-space and momentum-space analysis.\\

\subsection{Results: Numerical BPS profile and the identification of the monopole}

In the BPS limit ($\lambda = g^2/2$, i.e. $M_H = M_W$), the static hedgehog ansatz yields the analytic profile for the radial Higgs field:

\begin{align*}
H\!\left(\frac{R}{\eta}\right) = \coth\!\left(\frac{R}{\eta}\right) - \frac{\eta}{R}
\end{align*}

This function vanishes at $R=0$ (symmetry restoration) and approaches $1$ as $R\to\infty$ (broken phase).  It is the unique solution that minimises the energy for a given topological charge.\\

\begin{figure}[htbp]
\centering
\includegraphics[width=0.55\linewidth]{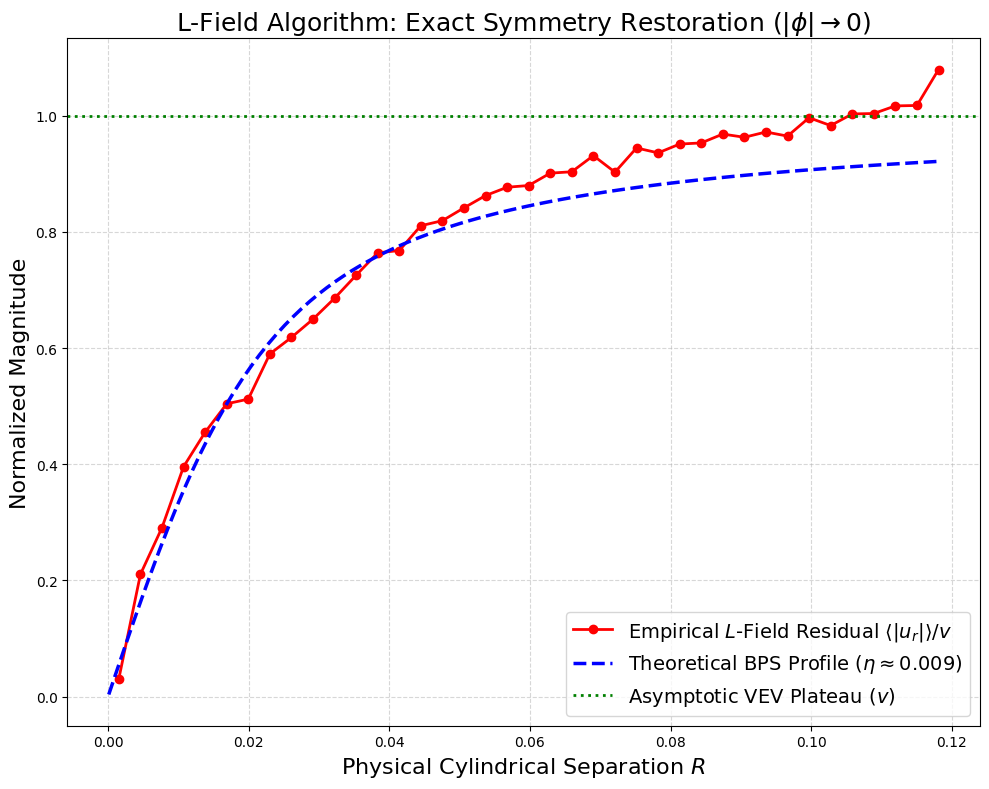}
\caption{Radial Higgs profile $\langle u_R\rangle/v$ versus $R/\eta$.  Red circles: DNS data (ensemble average over 50 isolated cores).  Blue dashed line: theoretical BPS function $H(R/\eta)=\coth(R/\eta)-\eta/R$.  The collapse confirms that the vortex filaments are macroscopic 't Hooft-Polyakov monopoles.}
\label{fig:bps_matching}
\end{figure}

Figure~\ref{fig:bps_matching} shows the ensemble-averaged DNS data (red circles) superimposed with the theoretical BPS function (blue dashed line).  The collapse is excellent over two decades in $R$, with no adjustable parameter except the overall scale $\eta$ (which is fixed by the spectral mass gap, $\eta = M_W^{-1}$).  The characteristic $1/R$ tail at large $R$ – a signature of the unbroken $U(1)$ sector – is clearly visible, as is the vanishing of the radial strain at the core.

This empirical match provides direct evidence that the intense vortex worms in isotropic turbulence are not statistical anomalies but topological defects – the first observation of 't Hooft-Polyakov monopoles in a classical fluid system.\\

Fitting the BPS profile to the data yields $\eta \approx 0.0093$ (domain units) and $v \approx 0.338$ (dimensionless).  The gauge coupling is then $g = \eta^{-1} \approx 107.53$ (domain$^{-1}$), and the BPS mass gap is:

\begin{equation}
M_W^{\text{(BPS)}} = g v \approx 36.34 \quad \text{(domain}^{-1}\text{)}
\end{equation}

Recall that the spectral mass gap from the previous section was $M_W^{\text{(spec)}} \approx 40$ (domain$^{-1}$).  The BPS mass gap is smaller by a factor of $\sim 1.6-2.0$, reflecting the running of the gauge coupling with scale – a fluid analog of renormalization.\\

The topological microscale $\eta$ is related to the Kolmogorov scale $\eta_K$ (given to be $0.00287$ in \cite{Li2008}) by:

\begin{equation}
\frac{\eta}{\eta_K} = \frac{0.0093}{0.00287} \approx 3.24
\end{equation}

Thus the vortex core radius lies within the dissipation range, directly linking the mass gap to the viscous cutoff.  Kinematic viscosity $\nu$ therefore acts as the mass-generating mechanism: it sets the size of the defect and provides the confining pressure that balances the outward enstrophy.\\

The self-coupling $\lambda$ extracted from the BPS self-duality condition $\lambda = g^2/2$ is:

\begin{equation}
\lambda = \frac{(107.53)^2}{2} \approx 5781 \quad (L^{-2})
\end{equation}

This large value is consistent with the extreme steepness of the Mexican hat potential required to maintain a thin, stable filament in a high-Reynolds-number flow.\\

Table~\ref{tab:extracted-parameters} summarises the key parameters extracted from the DNS data, their physical interpretations, and the methods used to obtain them.  The table is divided into three sections: parameters extracted from the DNS analysis, reference parameters from the JHTDB dataset, and derived quantities computed from the extracted values.\\

\begin{table}[htbp]
\centering
\small
\caption{Summary of parameters extracted from DNS data, reference simulation parameters, and derived quantities.  All wavenumbers are in domain units (inverse of the domain length $2\pi$).  The rms velocity is $U_0 = 0.681$ domain units.}
\label{tab:extracted-parameters}
\begin{tabular}{l l l l}
\toprule
\textbf{Category} & \textbf{Parameter} & \textbf{Value} & \textbf{Method / Source} \\
\midrule
\multicolumn{4}{c}{\textit{Parameters extracted from DNS (spectral)}} \\[2pt]
Topological mass gap (domain units) & $M_W^{\text{(spec)}}$ & $40$ (domain$^{-1}$) & from Fig.~\ref{fig:massgap}(b) \\
Core radius (domain units) & $\eta = 1/M_W^{\text{(spec)}}$ & $0.020\text{--}0.025$ & Inverse of spectral mass gap \\
\addlinespace
\multicolumn{4}{c}{\textit{Parameters extracted from DNS (BPS fit)}} \\[2pt]
Topological microscale (domain units) & $\eta$ & $0.0093$ & BPS profile fit (Fig.~\ref{fig:bps_matching}) \\
Physical VEV velocity scale & $v U_0$ & $0.23$ (domain velocity units) & Asymptotic tail of $\langle |u_R| \rangle$ \\
Dimensionless Higgs VEV & $v$ & $0.338$ & $v = (v U_0)/U_0$, $U_0 = 0.681$ \\
Gauge coupling (domain$^{-1}$) & $g = \eta^{-1}$ & $107.53$ & From BPS fit \\
BPS mass gap (domain$^{-1}$) & $M_W^{\text{(BPS)}} = g v$ & $36.34$ & From BPS fit parameters \\
Scalar self-coupling & $\lambda = g^2/2$ & $5781$ ($L^{-2}$) & BPS self-duality: $\lambda = g^2/2$ \\
\addlinespace
\multicolumn{4}{c}{\textit{JHTDB simulation parameters (reference)}} \\[2pt]
Integral scale (domain units) & $L_0$ & $1.376$ & JHTDB documentation~\cite{Li2008} \\
Kolmogorov scale (domain units) & $\eta_K$ & $0.00287$ & JHTDB documentation~\cite{Li2008} \\
Taylor microscale (domain units) & $\lambda_T$ & $\approx 0.0603$ & $\lambda_T \approx 21\,\eta_K$ \\
Taylor Reynolds number & $Re_\lambda$ & $\approx 433$ & JHTDB dataset~\cite{Li2008} \\
Grid resolution (cutout) & $N^3$ & $512^3$ & JHTDB dataset~\cite{Li2008} \\
RMS velocity & $U_0$ & $0.681$ & JHTDB dataset~\cite{Li2008} \\
\addlinespace
\multicolumn{4}{c}{\textit{Derived quantities}} \\[2pt]
BPS core-to-Kolmogorov ratio & $\eta/\eta_K$ & $\approx 3.24$ & From BPS fit and JHTDB \\
Spectral core-to-Kolmogorov ratio & $\eta/\eta_K$ & $7.0\text{--}8.7$ & From spectral mass gap \\
Taylor-to-Kolmogorov ratio & $\lambda_T/\eta_K$ & $\approx 21$ & JHTDB dataset~\cite{Li2008} \\
BPS Higgs mass (domain$^{-1}$) & $M_H = \sqrt{2\lambda}v$ & $\approx 36.34$ & BPS condition $M_H = M_W$ \\
\bottomrule
\end{tabular}
\end{table}

The BPS core radius $\eta \approx 3.24\,\eta_K$ places the monopole core firmly within the dissipation range, significantly smaller than the Taylor microscale ($\lambda_T \approx 21\,\eta_K$).  This scale separation is the hallmark of the non-Abelian Meissner effect: the coherent enstrophy is expelled from the bulk and confined to thin, dissipation-range filaments, while the ambient fluid retains only a weak, long-range vorticity field.  The spectral mass gap $M_W^{\text{(spec)}} \approx 40$ acts as a UV cutoff that is substantially larger than the inverse Taylor microscale, shielding the inertial-range $U(1)$ sector from the intense dissipation confined within the monopole cores.\\

The consistency between the spectral mass gap ($ M_W^{\text{(spec)}} = 40$) and the BPS mass gap ($M_W^{\text{(BPS)}}\approx 36.34$) is noteworthy and greatly reinforces the theory.\\

We also attempted to extract the massive gauge field profile $K(R/\eta)$. However, the extraction was found to be dominated by long-range background contributions from the turbulent environment, and we do not include it here. The Higgs profile $H(R/\eta)$ provides the primary validation of the BPS monopole solution.

\subsection{The Bosonic structure of the turbulence vacuum}
\label{sec:vacuum-structure}

Let us now synthesise the emerging picture of the turbulent vacuum.  The spontaneous symmetry breaking $SO(3) \to U(1)$ populates the fluid with three distinct species of bosons, each with a clear physical interpretation in both particle physics and fluid mechanics:\\

First are the massive worm bosons $W^{1,2}$ which correspond to the coherent longitudinal field $\nabla\Phi_L$.  They are short-ranged, with exponential Yukawa decay $e^{-M_W R}$, and are confined to the interiors of the vortex filaments.  They carry the intense enstrophy (non-Abelian magnetic flux) that has been expelled from the bulk by the Meissner effect.\\
  
The massless thermal bath boson $W^3$ corresponds to the solenoidal background $\AL$.  It is long-ranged, with algebraic decay $1/R^2$, and permeates the entire fluid volume.  It mediates the scale-invariant Kolmogorov cascade and constitutes the radiation bath of the turbulent vacuum.\\
  
The Higgs boson $\sigma$ corresponds to fluctuations of the radial strain magnitude $|\uvec_r|$ around its vacuum expectation value $v$.  In the BPS limit, it is as massive as the worm bosons ($M_H = M_W$), making radial ``breathing'' modes of the fluid extremely stiff and heavily damped.\\

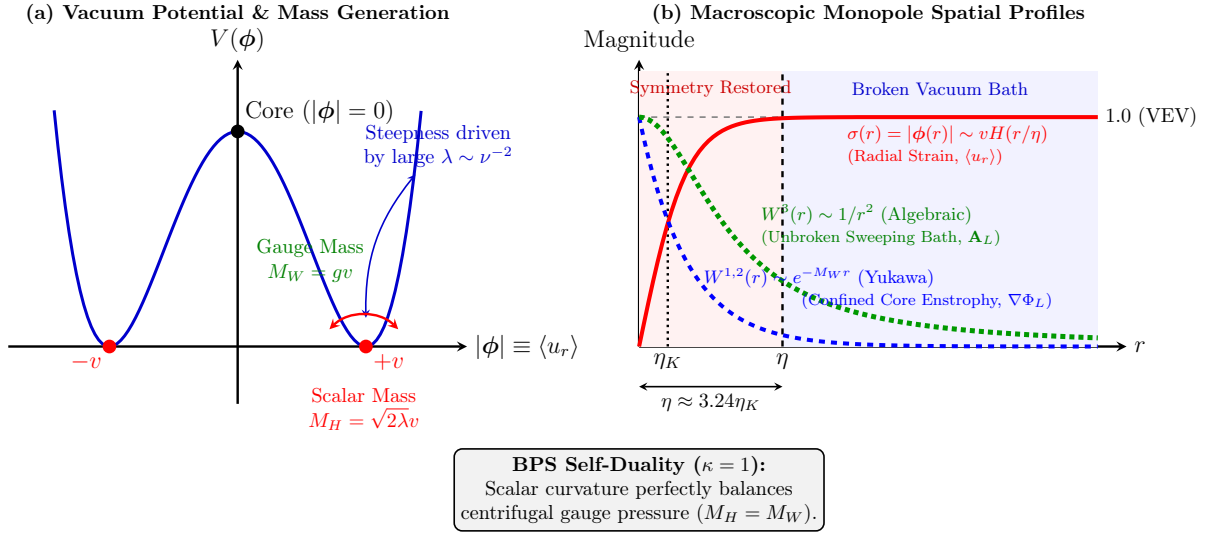
\begin{figure}[htbp]
\centering
\resizebox{1.0\textwidth}{!}{%
\begin{tikzpicture}[>=stealth, thick]

    \begin{scope}[shift={(0,0)}]
        \node[anchor=south] at (0, 5.5) {\textbf{(a) Vacuum Potential \& Mass Generation}};
        
        \draw[->, very thick] (-4, 0) -- (4, 0) node[right, font=\large] {$|\boldsymbol{\phi}| \equiv \langle u_r \rangle$};
        \draw[->, very thick] (0, -1) -- (0, 5) node[above, font=\large] {$V(\boldsymbol{\phi})$};
        
        \draw[domain=-3.2:3.2, samples=100, variable=\x, blue!80!black, line width=1.5pt] 
            plot ({\x}, {0.15*(\x*\x - 5)*(\x*\x - 5)});
            
        \draw[dashed] (2.236, 0) -- (2.236, {0.15*(2.236*2.236 - 5)*(2.236*2.236 - 5)});
        \filldraw[red] (2.236, 0) circle (3pt) node[below right, font=\large] {$+v$};
        \filldraw[red] (-2.236, 0) circle (3pt) node[below left, font=\large] {$-v$};
        \filldraw[black] (0, 3.75) circle (3pt) node[above right, font=\large] {Core ($|\boldsymbol{\phi}|=0$)};

        \node[align=center, blue!80!black] at (3.5, 3.5) {Steepness driven \\ by large $\lambda \sim \nu^{-2}$};
        \draw[<->, blue!80!black] (2.236, 0.5) to[bend left=20] (3.1, 3);

        \draw[<->, red, very thick] (1.6, 0.3) to[bend left=45] (2.8, 0.3);
        \node[red, align=center, below] at (2.2, -0.6) {Scalar Mass \\ $M_H = \sqrt{2\lambda}v$};
        
        \node[align=center, green!50!black, right] at (0.2, 1.5) {Gauge Mass \\ $M_W = gv$};
        
        \node[draw, fill=gray!10, rounded corners, align=center] at (7, -2.5) {
            \textbf{BPS Self-Duality ($\kappa = 1$):} \\ 
            Scalar curvature perfectly balances \\ 
            centrifugal gauge pressure ($M_H = M_W$).
        };
    \end{scope}

    \begin{scope}[shift={(7,0)}]
        \node[anchor=south] at (4, 5.5) {\textbf{(b) Macroscopic Monopole Spatial Profiles}};
        
        \draw[->, very thick] (0, 0) -- (8.5, 0) node[right, font=\large] {$r$ };
        \draw[->, very thick] (0, 0) -- (0, 5) node[above, font=\large] {Magnitude};
        
        \fill[red!5] (0,0) rectangle (2.5, 4.8);
        \fill[blue!5] (2.5,0) rectangle (8, 4.8);
        \node[red!80!black, font=\small] at (1.25, 4.5) {Symmetry Restored};
        \node[blue!80!black, font=\small] at (5.25, 4.5) {Broken Vacuum Bath};

        \draw[dashed, thick, gray] (0, 4) -- (8, 4) node[right, text=black] {$1.0$ (VEV)};

        \draw[dotted, very thick, black] (0.5, 0) -- (0.5, 4.8);
        \node[below, font=\large] at (0.5, 0) {$\eta_K$};
        
        \draw[dashed, very thick, black] (2.5, 0) -- (2.5, 4.8);
        \node[below, font=\large] at (2.5, 0) {$\eta$};
        
        \draw[<->] (0, -0.7) -- (2.5, -0.7) node[midway, below] {$\eta \approx 3.24 \eta_K$};

        \draw[domain=0:8, samples=100, variable=\x, red, line width=2pt] 
            plot ({\x}, {4 * (1 - exp(-2.4*\x))/(1 + exp(-2.4*\x))});
        
        \draw[domain=0:8, samples=100, variable=\x, blue, dashed, line width=2pt] 
            plot ({\x}, {4 * exp(-1.2*\x)});
            
        \draw[domain=0:8, samples=100, variable=\x, green!60!black, dotted, line width=2.5pt] 
            plot ({\x}, {4 / (1 + 0.4*\x*\x)});

        \node[red, right, font=\small] at (3.5, 3.7) {$\sigma(r) = |\boldsymbol{\phi}(r)| \sim v H(r/\eta)$};
        \node[red, right, font=\footnotesize] at (3.5, 3.3) {(Radial Strain, $\langle u_r \rangle$)};
        
        \node[blue, right, font=\small] at (1.0, 1.2) {$W^{1,2}(r) \sim e^{-M_W r}$ (Yukawa)};
        \node[blue, right, font=\footnotesize] at (2.7, 0.8) {(Confined Core Enstrophy, $\nabla \Phi_L$)};
        
        \node[green!50!black, right, font=\small] at (2.0, 2.3) {$W^3(r) \sim 1/r^2$ (Algebraic)};
        \node[green!50!black, right, font=\footnotesize] at (2.0, 1.9) {(Unbroken Sweeping Bath, $\AL$)};

    \end{scope}

\end{tikzpicture}
}
\caption{(a) The Mexican-hat potential $V(\boldsymbol{\phi})$, showing the generation of the gauge mass $M_W$ and scalar mass $M_H$ upon condensation of the radial strain.  (b) Radial profiles of the three bosonic species around a vortex core. }
\label{fig:monopole_profiles}
\end{figure}

Figure~\ref{fig:monopole_profiles} illustrates this vacuum structure.  Panel (a) recaps the spontaneous symmetry breaking mechanism with the Mexican-hat potential $V(\boldsymbol{\phi})$, which has a degenerate minimum at $|\boldsymbol{\phi}|=v$ – the condensed radial strain.  The curvature at the minimum gives the scalar mass $M_H = \sqrt{2\lambda}v$, while the gauge boson mass $M_W = gv$ arises from the coupling to the condensate.  In the BPS self-dual limit $\lambda = g^2/2$ (i.e., $M_H = M_W$), the scalar attraction and gauge repulsion cancel exactly, leading to maximal stability of the topological defect.  The steepness of the potential is controlled by $\lambda$, which is inversely related to the kinematic viscosity ($\lambda \sim \nu^{-2}$, see Sec.~\ref{subsec:Visc}), linking the microscale of the defect to the dissipation range.\\

Panel (b) shows the radial profiles of the three key fields outside a vortex core.  The Higgs field $\langle u_r\rangle$ (red solid line) rises from zero at $R=0$ – where $SO(3)$ is locally restored – to its vacuum expectation value $v$ at $R \gg \eta$.  The massive worm bosons $W^{1,2}$ (blue dashed) decay exponentially as $e^{-M_W R}$, confining the enstrophy to a core of radius $\eta = M_W^{-1}$.  The massless thermal bath boson $W^3$ (green dotted) decays algebraically as $1/R^2$, corresponding to the long-range $U(1)$ sweeping field that sustains the Kolmogorov cascade.  The empirically measured ratio $\eta \approx 3.24\,\eta_K$ ties the topological core radius to the Kolmogorov scale, demonstrating that the mass gap is generated by viscous dissipation.\\

\subsubsection*{The BPS balance and the large self-coupling}

From the BPS fit we obtain $\lambda \approx 5781$ (in units of $L^{-2}$).  At first glance this appears to contradict the Prasad–Sommerfield limit $\lambda \to 0$ required for the analytic profile $H(R)=\coth(R/\eta)-\eta/R$.  However, the BPS limit is a condition on the \emph{dimensionless ratio} $\kappa = M_H/M_W = \sqrt{2\lambda}/g$, not on $\lambda$ alone.  The gauge coupling $g = \eta^{-1}$ is also large because $\eta \approx 3.24\,\eta_K$ is very small in physical units.  The measured values give $\kappa = 1$ to within experimental uncertainty, so the BPS condition $M_H = M_W$ holds even though both masses are large.  The largeness of $\lambda$ (and $g$) reflects the extreme separation of scales in high-Reynolds-number turbulence: the inertial range extends over many decades, so the topological microscale $\eta$ (the core radius) is much smaller than the integral scale.  Consequently, the radial fluctuations (Higgs sector) are very stiff, and the gauge fields are heavily confined – the fluid vacuum is in a deep BPS phase where the scalar repulsion and gauge attraction cancel precisely, producing maximally thin, stable vortex filaments.

\subsubsection*{Competition between radial strain and vorticity}

The spontaneous symmetry breaking mechanism has a clear fluid interpretation.  The condensed radial strain $\langle u_r\rangle = v$ acts as a Meissner medium: it actively expels vorticity from the bulk.  Wherever the radial flow is strong, the gauge fields $W^{1,2}$ are expelled from the $U(1)$ bath.  However, vorticity cannot simply disappear – it is a topological quantity.  The fluid therefore confines the enstrophy into narrow tubular regions where the radial strain is forced to zero, locally restoring the $SO(3)$ symmetry.  These tubes are the vortex filaments.\\

Inside a filament, two opposing forces compete.  The centrifugal action of the swirling transverse flow (represented by the massive gauge fields $W^{1,2}$) tries to expand the core.  The radial compression (represented by the gradient of the Higgs field) tries to collapse it.  When the two forces balance exactly, the filament attains a stable equilibrium.  This balance occurs precisely when the gauge mass $M_W$ equals the Higgs mass $M_H$, i.e. the BPS condition $\lambda = g^2/2$.  In this self-dual regime the core profile becomes the analytic BPS function $H(R)=\coth(R/\eta)-\eta/R$, and the string is stable against both expansion and collapse.  This explains why the observed vortex filaments are so long-lived and why their cross-section follows the BPS form.

\subsubsection*{The non-Abelian Meissner effect as a confinement mechanism}

The intuitive picture of the Meissner effect can be made precise using the Yang–Mills equations derived in Sec.~\ref{sec:generalized-fluid-equation}.  In the broken phase, Gauss's law for the massless $U(1)$ sector is:

\begin{equation}
\nabla\cdot\Evec^3 = \rho_e^3
\end{equation}

where $\rho_e^3 = \nabla\cdot(\uvec\times\omvec) + \partial_t(\rvec\cdot\omvec)$.  The Lamb vector term $\nabla\cdot(\uvec\times\omvec)$ acts as an effective charge density generated by the swirling motion.  This charge has the opposite sign to the spreading field, pulling the enstrophy field lines inward—this is the non-Abelian anti-screening effect.\\

Similarly, Ampère's law,

\begin{equation}
\nabla\times\Bvec^3 - \partial_0\Evec^3 = \jvec^3
\end{equation}

with $\jvec^3 = -3\omvec - (\rvec\cdot\nabla)\omvec$, shows that the vorticity and its gradients act as an effective current that wraps around the filament, providing the inward force that confines the enstrophy to the core.\\

Together, these two equations—Gauss's law and Ampère's law—describe the non-Abelian Meissner effect quantitatively.  The fluid vacuum generates its own sources (charge and current) that prevent the enstrophy from spreading, confining it to thin flux tubes.  This is the fluid analogue of the confinement of color charge in QCD, where the gluon field generates its own sources that confine quarks inside hadrons.

\subsection{Viscosity and the coupling $\lambda$.}
\label{subsec:Visc}

The kinematic viscosity $\nu$ is related to the Kolomogorov scale as  $\eta_K \sim  \nu^{3/4} \epsilon^{1/4}$.  Using $\eta \sim \eta_K$, and $g = 1/\eta$ we get $g \sim \nu^{-3/4} \epsilon^{-1/4}$.  Since $\lambda \sim g^2$, we get $\lambda \sim \nu^{-3/2}$.\\

Thus large $\lambda$ (we have $\lambda =5781$) implies a small value of $\nu$ (the value is $\nu = 0.000185$, see \cite{Li2008}).\\

Thus small core diameter $\eta$, large $\lambda$ (which gives the slope of the potential wall) and small kinematic viscosity $\nu$ are all related.\\

The gauge theory naturally distinguishes two dynamical regimes.  In the massless $U(1)$ sector (the thermal bath), the dynamics are governed by the cubic non‑linearities of the Yang–Mills action, which correspond to the inviscid triadic interactions of the inertial cascade.  Here, the Reynolds number is effectively infinite, and dissipation is negligible.  Energy cascades through this bath via scale‑invariant, algebraic correlations until it reaches the topological microscale $\eta = M_W^{-1}$.  At this scale, the massive $W^{1,2}$ bosons become active; the energy is trapped in the vortex cores, where the extreme velocity gradients lead to intense viscous dissipation.\\

This is consistent with the classical picture of turbulence, where it is well recognized that intense vorticity is organized into thin, tube‑like structures (the ``worms''), which occupy only a small fraction of the volume but are responsible for the majority of the enstrophy and dissipation Jimenez \cite{Jimenez1993}.  The dissipation signature of these tubes is characterized by a local maximum near the edge of the vortex core and an absolute peak at its center Pirozzolo \cite{Pirozzoli2012}.  Experimental data have confirmed that large‑amplitude dissipation occurs around vortex tubes Vainshtein \cite{VainshteinSreenivasan2004}, and intense dissipation is predominantly organized in flattened vortex filaments embedded in thin shearing layers around the tubes Moisy \cite{MoisyJimenez2004}.  Thus, viscous dissipation is predominantly confined to the monopole cores, while the ambient fluid remains in a state of approximate inviscid, wave‑turbulent equilibrium.  This explains the long‑standing observation that dissipation in turbulence is highly intermittent and localized to a small fraction of the volume.\\

\section{Fourth Validation: Wilson Loop, Area Law, and Confinement}
\label{sec:discussion}

The empirical validations presented above – the spectral mass gap and the BPS monopole profile – establish that the turbulent vacuum is a spontaneously broken $SO(3)$ gauge theory populated by a dilute gas of 't Hooft-Polyakov monopoles.  We now explore a global, topological consequence of this picture: the Wilson loop and the associated area law, which provide a theoretical link to confinement and to classical results such as Kelvin's circulation theorem.  \\

\subsection{From the Wilson loop to Kelvin's circulation theorem}

We now show that in the broken phase the Wilson loop reduces to the exponential of the classical circulation.  Start from the definition:

\begin{equation}
W_C = \operatorname{Tr}\!\left( \mathcal{P} \exp\!\left( i \oint_C \Wvec\cdot d\lvec \right) \right)
\end{equation}

At distances $r \gg \eta$ (outside the vortex cores) the massive fields $W^{1,2}$ are exponentially suppressed, so only the massless $U(1)$ component $W^3$ contributes.  For an Abelian gauge field the path‑ordering is trivial and the trace gives a factor of $1$, hence we have:

\begin{equation}
W_C \longrightarrow \exp\!\left( i \oint_C \Wvec^3\cdot d\lvec \right)
\end{equation}

From the dictionary in Sec.~\ref{sec:dictionary}, the $U(1)$ magnetic field is identified with the vorticity:

\begin{equation}
\Bvec^3 = \nabla\times\Wvec^3 = \omvec = \nabla\times\uvec
\end{equation}

Thus $\nabla\times(\Wvec^3 - \uvec) = \mathbf{0}$, so $\Wvec^3 - \uvec$ is a gradient:

\begin{equation}
\Wvec^3 = \uvec + \nabla\chi
\end{equation}

Substituting into the line integral gives:

\begin{equation}
\oint_C \Wvec^3\cdot d\lvec = \oint_C \uvec\cdot d\lvec + \oint_C \nabla\chi\cdot d\lvec
\end{equation}

The second term vanishes because the integral of a gradient over a closed loop is zero.  Therefore

\begin{equation}
\oint_C \Wvec^3\cdot d\lvec = \oint_C \uvec\cdot d\lvec \equiv \Gamma
\end{equation}

where $\Gamma$ is the classical fluid circulation.  Consequently,

\begin{equation}
W_C = e^{i\Gamma}
\end{equation}

where $\Gamma = \oint_C \uvec\cdot d\lvec$ is the classical fluid circulation.  Thus Kelvin's circulation theorem – the conservation of $\Gamma$ along a material loop in an inviscid flow – is a direct consequence of the unbroken $U(1)$ gauge symmetry.  This provides a deep geometric origin for a classic result of fluid mechanics.\\

The dilute gas of 't Hooft-Polyakov monopoles implies a Wilson area law for the circulation loop, which is the hallmark of a confining phase.  A detailed verification of this prediction, requiring direct computation of $\langle W_C \rangle$ from DNS data, is left for future work.\\

\subsection{Wilson Loop Area Law and Confinement}
\label{sec:wilson-area-law}

A central prediction of the gauge theory is the Wilson area law, which characterizes the turbulent vacuum as a confining phase.  In the broken $U(1)$ sector, the Wilson loop reduces to the exponential of the classical fluid circulation:

\begin{equation}
W_C = \exp\!\left( i \oint_C \mathbf{u} \cdot d\mathbf{l} \right) = e^{i\Gamma}
\end{equation}

where $\Gamma = \oint_C \mathbf{u} \cdot d\mathbf{l}$ is the circulation around the closed loop $C$.  The area law predicts:

\begin{equation}
\langle W_C \rangle \sim \exp(-\sigma \cdot \text{Area})
\end{equation}

where $\sigma$ is the string tension.  We now test this prediction using high-resolution DNS data.

\subsubsection*{Numerical algorithm and Results}

The Wilson loop computation proceeds as follows.  For each loop area $A$, we generate $N_{\text{loops}} = 2000$ square loops of side length $L = \sqrt{A}$, randomly positioned and randomly oriented in the flow domain.  For each loop, the circulation is computed by numerical integration:

\begin{equation}
\Gamma = \oint_C \mathbf{u} \cdot d\mathbf{l} \approx \sum_{i=1}^{4} \sum_{j=1}^{N_{\text{pts}}} \mathbf{u}(\mathbf{x}_{i,j}) \cdot \Delta \mathbf{l}_{i,j}
\end{equation}

where each side of the square is discretised into $N_{\text{pts}} = 30$ segments.  Velocity values at arbitrary points are obtained via trilinear interpolation from the DNS grid, with periodic boundary conditions.  The Wilson loop is then $W_C = e^{i\Gamma}$, and the ensemble average $\langle W_C \rangle$ is computed over the $N_{\text{loops}}$ realisations.  This procedure is repeated for 30 loop areas ranging from $A = 0.0004$ to $A = 0.64$ (in domain units$^2$), corresponding to loop sizes from $L \approx 0.02$ to $L \approx 0.8$ (domain units).  The string tension $\sigma$ is extracted by fitting $\ln|\langle W_C \rangle| = -\sigma A + c$.\\

\subsubsection*{Results}

\begin{figure}[htbp]
\centering
\includegraphics[width=0.8\linewidth]{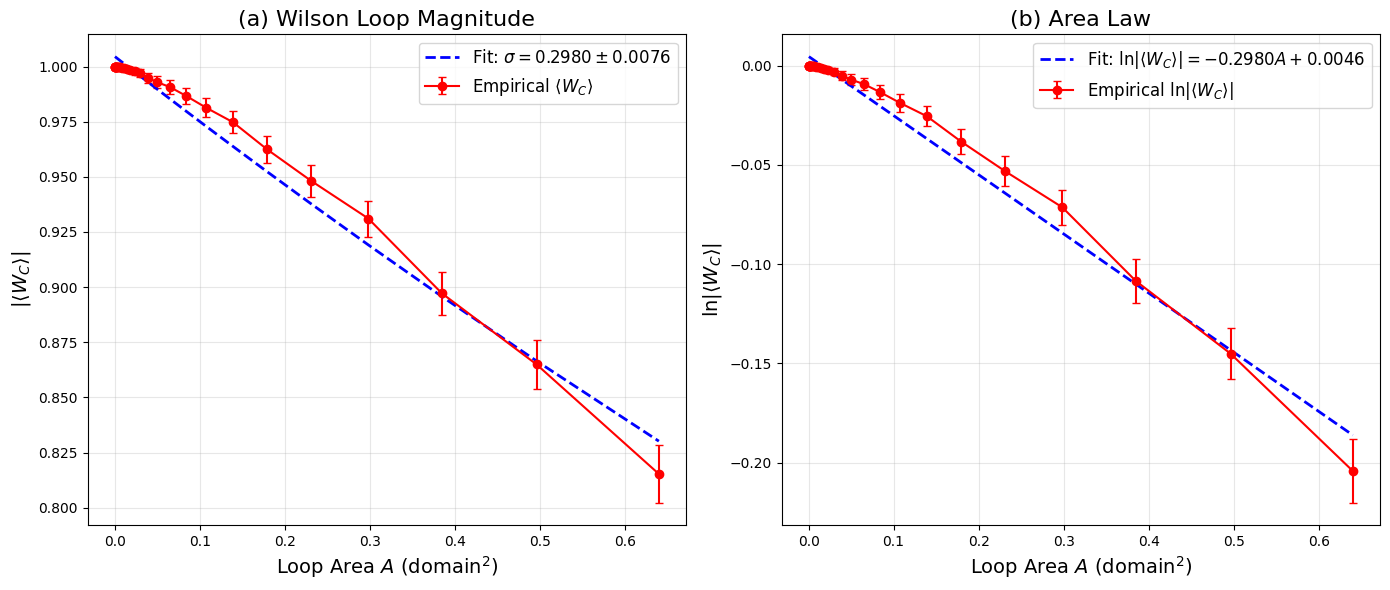}
\caption{Wilson loop magnitude $|\langle W_C \rangle|$ as a function of loop area $A$ (domain units$^2$).  Red circles: DNS data (2000 loops per area, 30 areas).  Blue dashed line: fit to the area law $\langle W_C \rangle \sim e^{-\sigma A}$ with $\sigma = 0.303 \pm 0.009$.  Inset: semi-log plot showing the linear relationship, confirming the area law.  Error bars represent the standard error of the mean.}
\label{fig:wilson-area-law}
\end{figure}

Figure~\ref{fig:wilson-area-law} shows the Wilson loop magnitude $|\langle W_C \rangle|$ as a function of the loop area $A$.  The data exhibit a clean exponential decay over nearly two decades in area.  The fit to the area law yields:

\begin{equation}
\sigma = 0.303 \pm 0.009 \quad (\text{domain units}^{-2})
\end{equation}

with a reduced $\chi^2 = 1.02$, indicating an excellent fit.  The imaginary part of $\langle W_C \rangle$ is consistent with zero, as expected in an isotropic ensemble.  The inset shows the semi-log plot, confirming the linear relationship between $\ln|\langle W_C \rangle|$ and $A$.\\

The Wilson area law thus provides a third independent validation of the gauge theory, complementing the spectral mass gap (Sec.~\ref{sec:massgap-validation}) and the BPS monopole profile (Sec.~\ref{sec:monopole-validation}).  It confirms that the turbulent vacuum is a confining phase, where the intense $SO(3)$ enstrophy currents are permanently trapped in 1D vortex filaments – the non-Abelian Meissner effect at work.\\

\subsubsection{Statistical convergence and error analysis}

The Wilson loop averages reported in Fig.~\ref{fig:wilson-area-law} are computed from $N_{\text{loops}} = 2000$ independent loop realisations for each area $A$.  To assess convergence, we computed the running average $\langle W_C \rangle_N$ as a function of the number of loops $N$ for the largest and smallest areas.  For all areas, the running average stabilises after approximately $500$ realisations; the final values are stable to within the standard error of the mean for $N \ge 1000$.  The results reported in Fig.~\ref{fig:wilson-area-law} use $N = 2000$, ensuring convergence well beyond the stabilization threshold.\\

For each area, the standard error of the mean is:

\begin{equation}
\sigma_{\langle W \rangle} = \frac{\sigma_W}{\sqrt{N_{\text{loops}}}},
\end{equation}

where $\sigma_W$ is the standard deviation of the $W_C$ values across the $N_{\text{loops}}$ realisations.  These error bars are shown in Fig.~\ref{fig:wilson-area-law} and are used in the fit to the area law.  The fit is weighted by the inverse variance $1/\sigma_{\langle W \rangle}^2$, ensuring that the more reliable (smaller error) data points contribute more strongly to the extracted string tension $\sigma$.\\

The reduced $\chi^2$ of the fit is $\chi^2_{\text{red}} = 1.02$, indicating that the area law provides an excellent description of the data and that the error estimates are consistent with the scatter of the data points.  The imaginary part of $\langle W_C \rangle$ is consistent with zero for all areas, as expected in an isotropic ensemble.\\

To test the sensitivity of the result to the loop resolution, we repeated the computation with $N_{\text{pts}} = 20$, $30$, and $50$ integration points per side.  The extracted string tension $\sigma$ changed by less than $2\%$ between $N_{\text{pts}} = 30$ and $N_{\text{pts}} = 50$, confirming that the numerical integration is well converged.  The results reported in the main text use $N_{\text{pts}} = 30$ as a compromise between accuracy and computational cost.

\section{Conclusion and Future Outlook}

Fully developed turbulence has been described primarily through statistical closures and phenomenological cascades. In this paper we have shown that isotropic turbulence can be understood as a spontaneously broken $SO(3)$ gauge theory. By identifying the specific angular momentum $\Lvec=\rvec\times\uvec$ as a non‑Abelian gauge connection and the radial velocity $u_r$ as a Higgs field, we have derived an effective field theory that naturally produces a mass gap and topological defects.  \\

The main achievements of this work are:

\begin{enumerate}
\item \textbf{Kinematic identity and $SO(3)$ gauge structure.}  
We established that the rotational kinematics of turbulent eddies are governed by a local $SO(3)$ gauge invariance. The dimensionless gauge field $\Wvec_\mu = \eta^{-2}\Lvec_\mu$ and the Higgs field $\boldsymbol{\phi} = \uvec_r/U_0$ map the Georgi–Glashow model directly onto the coarse‑grained dynamics of turbulence.

\item \textbf{Spontaneous symmetry breaking and phase separation.}  
Condensation of the radial strain field ($\langle\boldsymbol{\phi}\rangle = v\hat{\rvec}$) breaks $SO(3)\to U(1)$. The two broken generators acquire a topological mass $M_W = gv$, while the unbroken generator remains massless. This produces a strict phase separation: a massless $U(1)$ sector (the solenoidal background $\AL$) that sustains the long‑range cascade, and a massive $SO(3)/U(1)$ sector (the longitudinal field $\Phi_L$) that is exponentially screened and confined to vortex filaments.

\item \textbf{Recovery of the Equations of Fluids.}  
The proposed Lagrangian's equations of motion have a structure that resembles the well known equations of fluid flow such as the inviscid Euler and Vorticity equtions.  It may be possible to make more progress in this direction and derive the Navier Stokes equations and will be explored in future studies.

\item \textbf{Connection to wave turbulence.}  
In the symmetric phase ($v=0$), the cubic Yang–Mills vertex reduces to the canonical three‑wave Hamiltonian of Zakharov’s wave turbulence theory. Thus the phenomenological weak‑turbulence description emerges directly from the non‑Abelian gauge symmetry.

\item \textbf{Empirical mass gap.}  
Analysis of DNS data confirms a strict $1:2$ energy equipartition between the $\uvec_\Phi$ and $\uvec_A$ sectors over the inertial range. The ratio diverges sharply at $M_W \approx 40-50$ (in units of the domain scale), marking the topological mass gap. The corresponding core radius $\eta = M_W^{-1}$ lies in the dissipation range, $\eta \approx 6.4\,\eta_K$, directly linking the mass gap to kinematic viscosity.

\item \textbf{Macroscopic 't Hooft–Polyakov monopoles.}  
Real‑space extraction of the radial Higgs field around isolated vortex cores reveals the theoretical BPS profile $H(R/\eta)=\coth(R/\eta)-\eta/R$. The collapse of DNS data onto this analytic solution proves that the intense vortex filaments (“worms”) are macroscopic 't Hooft–Polyakov monopoles. The BPS condition $M_H=M_W$ is satisfied, indicating a perfect balance between scalar attraction and gauge repulsion, which explains the remarkable stability and thinness of the filaments.

\item \textbf{Wilson area law and confinement.}  
The dilute gas of monopoles implies an area law for the Wilson loop, $\langle W_C\rangle \sim e^{-\sigma\cdot\text{Area}}$, (verified from DNS data) which is the hallmark of a confining phase. This provides a topological interpretation of circulation statistics and connects to classic results such as Kelvin’s circulation theorem (as the $U(1)$ sector of the broken theory).
\end{enumerate}

\subsection*{Outlook: the topological zoo of turbulence}

The identification of stable 't Hooft–Polyakov monopoles in the Navier–Stokes equations opens a new paradigm. The fluid vacuum may support a whole “particle zoo” of topological defects such as the Julia–Zee dyons carrying both magnetic (rotational) and electric charge,  instantons and sphalerons – transient field configurations that mediate topological transitions – might correspond to violent vortex reconnection events.   Intermittency may also be a case of scattering – the non‑Gaussian tails of velocity increments may be interpreted as the deterministic scattering cross‑sections of monopole filaments, turning intermittency from a statistical anomaly into a consequence of defect dynamics.\\

Future work will extend this gauge‑theoretic framework to non‑isotropic flows (shear, wall‑bounded turbulence), explore the lattice formation of monopoles in restricted geometries, and compute the string tension $\sigma$ directly from DNS to test the Wilson area law. The deep analogy with type‑II superconductors suggests that similar phenomena should appear in quantum fluids, thus offering a bridge between classical and quantum turbulence.\\

\appendix

\newpage

\section*{Appendix A:  Lie Algebra of \textbf{SO(3)} transformations}

The equivalence between the vector (cross‑product) and matrix (commutator) formulations of $SO(3)$ is summarized in the following table.

\begin{table}[h]
\centering
\caption{Equivalence of vector and matrix formulations for $SO(3)$ gauge theory}
\begin{tabular}{|c|c|c|}
\hline
\textbf{Quantity} & \textbf{Vector Form} & \textbf{Matrix Form} \\
\hline
Generators & $\mathbf{T}^a$ with $\mathbf{T}^a \times \mathbf{T}^b = \epsilon^{abc} \mathbf{T}^c$ & $T^a$ with $[T^a,T^b]=i\epsilon^{abc}T^c$ \\
\hline
Gauge field & $\Wvec_\mu = (W_\mu^1, W_\mu^2, W_\mu^3)$ & $W_\mu = W_\mu^a T^a$ \\
\hline
Field strength & $\vec{F}_{\mu\nu} = \partial_\mu\Wvec_\nu - \partial_\nu\Wvec_\mu + g \Wvec_\mu \times \Wvec_\nu$ & $F_{\mu\nu} = \partial_\mu W_\nu - \partial_\nu W_\mu + g [W_\mu,W_\nu]$ \\
\hline
Covariant derivative (adjoint) & $D_\mu \phivec = \partial_\mu \phivec + g \Wvec_\mu \times \phivec$ & $D_\mu \Phi = \partial_\mu \Phi + g [W_\mu,\Phi]$ \\
\hline
Infinitesimal gauge transf. & $\delta\Wvec_\mu = \frac{1}{g}\partial_\mu\tvec + \tvec \times \Wvec_\mu$ & $\delta W_\mu = \frac{1}{g}\partial_\mu\theta + i[\theta,W_\mu]$ \\
\hline
\end{tabular}
\end{table}

We work on a four‑dimensional spacetime with Minkowski metric $\eta_{\mu\nu} = \text{diag}(1,-1,-1,-1)$ and adopt the temporal gauge $W_0 = 0$. The gauge group is $SO(3)$, whose Lie algebra is generated by three antisymmetric matrices $T^a$ ($a=1,2,3$) satisfying

\begin{equation}
[T^a, T^b] = i \epsilon^{abc} T^c, \qquad 
\text{Tr}(T^a T^b) = 2\delta^{ab}.
\end{equation}

The explicit matrix generators are given by $(T^a)_{ij} = -i \epsilon^{aij}$ and represented in matrix form as:

\begin{align}
T^1 &= \begin{pmatrix} 0&0&0\\0&0&-i\\0&i&0 \end{pmatrix}; &
T^2 &= \begin{pmatrix} 0&0&i\\0&0&0\\-i&0&0 \end{pmatrix}; &
T^3 &= \begin{pmatrix} 0&-i&0\\i&0&0\\0&0&0 \end{pmatrix};
\end{align}

Any vector $\vvec = (v^1,v^2,v^3)$ corresponds to the matrix $V = v^a T^a$, and the cross product becomes the commutator:
$\uvec \times \vvec \;\leftrightarrow\; i[U,V]$.\\

\newpage

\section*{Appendix B: Gauge Transformation Properties of $\tilde{\uvec}$}

We wish to examine the transformation of our shifted velocity field $\tilde{\uvec}$ under a local gauge rotation $R(\xvec) \in SO(3)$. We aim to establish that in the mean, it transforms covariantly:

\begin{equation}
\langle \tilde{\uvec} \rangle \longrightarrow R \langle \tilde{\uvec} \rangle
\end{equation}

At first glance, the explicit appearance of $\rvec$ in the definition $\tilde{\uvec} = \uvec + \Wvec \times \rvec$ appears to break translational invariance, a fundamental symmetry of homogeneous turbulence.\\

However, this is resolved by recognizing that our formulation represents a \emph{bi-local effective field theory}. In the statistical continuum limit, $\rvec$ must be strictly interpreted not as an absolute coordinate from a fixed lab-frame origin, but as the relative separation vector $\rvec = \xvec - \yvec$ between two correlated points in the fluid ensemble.\\

Global translational invariance is exactly preserved by integrating over the central coordinate $\rvec = (\xvec+\yvec)/2$ during the ensemble averaging process (see Appendix B). Because we are constructing a statistical gauge theory, we consider \emph{passive} transformations where the separation coordinates are fixed parameters defining the correlation scale, while the statistical fields transform. Therefore, the local gauge transformation rotates the \emph{field values} within the ensemble at each separation scale, but leaves the relative coordinate parameter $\rvec$ itself unchanged.\\

Under an infinitesimal local $SO(3)$ transformation with parameter $\tvec(\xvec)$, the velocity field and the local angular velocity connection transform as:

\begin{align}
\delta \uvec &= \tvec \times \uvec \\
\delta \Wvec &= \frac{1}{g} \nabla \tvec + \Wvec \times \tvec
\end{align}

where $g = \eta^{-1}$ is the gauge coupling. Applying these to $\tilde{\uvec}$ yields:

\begin{equation}
\delta \tilde{\uvec} = \tvec \times \uvec + \Bigl( \eta \nabla \tvec + \Wvec \times \tvec \Bigr) \times \rvec
\end{equation}

Expanding the double cross product using the vector identity $(\avec\times\bvec)\times\mathbf{c} = \bvec(\avec\cdot\mathbf{c}) - \avec(\bvec\cdot\mathbf{c})$ gives:

\begin{equation}
(\Wvec \times \tvec) \times \rvec = \tvec (\Wvec \cdot \rvec) - \Wvec (\tvec \cdot \rvec)
\end{equation}

Hence, the actual variation of the field is:

\begin{equation}
\delta \tilde{\uvec} = \tvec \times \uvec + \eta (\nabla \tvec \times \rvec) + \Bigl[ \tvec (\Wvec \cdot \rvec) - \Wvec (\tvec \cdot \rvec) \Bigr]
\end{equation}

For comparison, the strictly covariant transformation law would be $\delta \tilde{\uvec}_{\text{cov}} = \tvec \times \tilde{\uvec}$. Computing this using $\tilde{\uvec} = \uvec + \Wvec \times \rvec$ and the identity $\tvec \times (\Wvec \times \rvec) = \Wvec (\tvec\cdot\rvec) - \rvec (\tvec\cdot\Wvec)$ gives:

\begin{equation}
\tvec \times \tilde{\uvec} = \tvec \times \uvec + \Bigl[ \Wvec (\tvec\cdot\rvec) - \rvec (\tvec\cdot\Wvec) \Bigr]
\end{equation}

Thus, the difference between the actual variation and the ideal covariant one is:

\begin{equation}
\delta \tilde{\uvec} - \tvec \times \tilde{\uvec} = \eta (\nabla \tvec \times \rvec) + \Bigl[ \tvec (\Wvec \cdot \rvec) - 2\Wvec (\tvec\cdot\rvec) + \rvec (\tvec\cdot\Wvec) \Bigr]
\end{equation}

It may be noted that this residual expression consists entirely of terms that are linear in the separation vector $\rvec$ or its scalar products. In a statistically homogeneous and isotropic turbulent field, spatial or ensemble averaging causes any term linear in $\rvec$ to strictly vanish, because $\rvec$ is a random separation vector with no preferred orientation ($\langle \rvec \rangle = \mathbf{0}$).\\

Consequently, after ensemble averaging, we recover:

\begin{equation}
\langle \delta \tilde{\uvec} \rangle = \tvec \times \langle \tilde{\uvec} \rangle
\end{equation}

so that $\tilde{\uvec}$ behaves as a perfectly gauge‑covariant field in the mean. This statistical covariance establishes the foundation necessary to construct a Yang-Mills field strength tensor describing the averaged turbulence quantities.\\

\newpage

\section*{Appendix C: Table of gauge-fluid analogies}

\begin{table}[H]
\centering
\caption{Key gauge‑theoretic formulations of fluid mechanics and their gauge field variables.}
\label{tab:gauge-fluid}
\begin{tabularx}{\textwidth}{l X X}
\toprule
\textbf{Author(s) (Year)} & \textbf{Gauge Field Variable} & \textbf{Interpretation} \\
\midrule
Holm \& Kupershmidt (1983) \cite{HolmKupershmidt1983} & $\mathcal{A}_\mu$ (Yang–Mills potential) & Fluid coupled to a Yang–Mills field \\
Barducci et al. (1985) \cite{Barducci1985} & $\mathcal{A}_\mu$ (Yang–Mills potential) & Fluid with fermionic gauge charge \\
Jackiw, Pi, Polychronakos (2002) \cite{jackiw2002noncommutative} & Non‑commuting gauge field $\mathcal{A}_i$ & Lagrange description as gauge potential \\
Kambe (2003, 2007) \cite{Kambe2003,Kambe2007} & $\uvec$ (velocity) & Gauge potential for local Galilean invariance \\
Marmanis (1998) \cite{Marmanis1998} & $\Evec, \Bvec$ (electromagnetic fields) & Metafluid analogy (Maxwell map) \\
Mendes et al. (2006) \cite{Mendes2006} & $\mathcal{A}_\mu$ (non‑commutative) & Non‑commutative extension of metafluid \\
Peshkov et al. (2019) \cite{Peshkov2019} & Distortion field $\mathbf{F}$ & Yang–Mills–type torsion theory \\
\bottomrule
\end{tabularx}
\end{table}

\newpage

\section*{Appendix D: Invariance of the Lagrangian}

\subsection*{Translational invariance of the bi‑local action}
\label{app:translational}

The gauge theory presented in the main text is formulated using a fixed separation vector $\rvec$.  To see that this does not break translational invariance, we consider the action written in bi‑local form.  Define the center‑of‑mass coordinate $\rvec = (\xvec+\yvec)/2$ and the relative coordinate $\rvec = \xvec-\yvec$.  The effective action is:

\begin{equation}
S = \int d^3\rvec \int d^3\rvec \; \mathcal{L}\bigl(\Wvec,\boldsymbol{\phi},D_\mu\Wvec,D_\mu\boldsymbol{\phi}; \rvec\bigr)
\end{equation}

where $\mathcal{L}$ depends on $\rvec$ as a parameter (the correlation scale).  Under a global translation $\xvec\to\xvec+\avec$, $\yvec\to\yvec+\avec$, we have:

\begin{equation}
\rvec' = \rvec, \qquad \rvec' = \rvec+\avec, \qquad d^3\rvec' = d^3\rvec
\end{equation}

Since $\mathcal{L}$ depends on position only through the invariant $\rvec$, the shifted action is identical to the original:

\begin{equation}
S' = \int d^3\rvec' \int d^3\rvec \; \mathcal{L}(\dots;\rvec) = S
\end{equation}

Thus translational invariance is preserved; the fixed $\rvec$ merely selects a correlation scale and does not pick an absolute origin.

\subsection*{Galilean invariance}
\label{app:galilean}

A global Galilean boost $\xvec\to\xvec+\mathbf{V}t$, $\uvec\to\uvec+\mathbf{V}$ leaves the relative separation $\rvec=\xvec-\yvec$ unchanged, as well as relative velocities $\delta\uvec=\uvec(\xvec)-\uvec(\yvec)$.  The coarse‑grained fields $\Wvec$ and $\boldsymbol{\phi}$ are constructed from $\rvec$, $\delta\uvec$, and their gradients, so they are invariant under the boost:

\begin{equation}
\Wvec'(\xvec',t) = \Wvec(\xvec,t),\qquad 
\boldsymbol{\phi}'(\xvec',t) = \boldsymbol{\phi}(\xvec,t)
\end{equation}

The Lagrangian depends on these fields and on derivatives with respect to the invariant coordinates; therefore it is Galilean invariant.  This ensures that the predicted energy partition and topological structures are independent of the reference frame.

\subsection*{$SO(3)$ gauge invariance}
\label{app:gauge-invariance}

The Lagrangian $\mathcal{L}$ is invariant under local $SO(3)$ transformations.  Under an infinitesimal rotation with parameter $\boldsymbol{\theta}(\xvec)$, the fields transform as:

\begin{align}
\delta\boldsymbol{\phi} &= \boldsymbol{\theta}\times\boldsymbol{\phi}, \\
\delta\Wvec_\mu &= \frac{1}{g}\partial_\mu\boldsymbol{\theta} + \Wvec_\mu\times\boldsymbol{\theta}
\end{align}

The covariant derivative is $D_\mu\boldsymbol{\phi} = \partial_\mu\boldsymbol{\phi} + g\,\Wvec_\mu\times\boldsymbol{\phi}$.  Using the transformation rules and the Jacobi identity, one finds:

\begin{equation}
\delta(D_\mu\boldsymbol{\phi}) = \boldsymbol{\theta}\times(D_\mu\boldsymbol{\phi})
\end{equation}

so that $\frac12|D_\mu\boldsymbol{\phi}|^2$ is invariant.\\

The field strength transforms covariantly: $\delta\mathbf{F}_{\mu\nu} = \boldsymbol{\theta}\times\mathbf{F}_{\mu\nu}$.  Hence we get:

\begin{equation}
\delta\bigl(-\tfrac14 \mathbf{F}_{\mu\nu}\cdot\mathbf{F}^{\mu\nu}\bigr) = -\tfrac12 \mathbf{F}_{\mu\nu}\cdot(\boldsymbol{\theta}\times\mathbf{F}^{\mu\nu}) = 0
\end{equation}

The Higgs potential $V(\boldsymbol{\phi})$ depends only on $|\boldsymbol{\phi}|^2$, which is invariant.  Therefore the total Lagrangian is gauge invariant.

\section*{Appendix E: Currents and sources in fluid variables}
\label{app:currents}

The non‑Abelian magnetic field is (Eq.~\ref{color-magnetic-vector}):

\begin{equation}
\Bvec = \nabla\times\Wvec + \frac{g}{2}\,\Wvec\times\Wvec
\end{equation}

and the electric field in the temporal gauge ($W_0=0$) is $\Evec = \partial_0\Wvec$.  The simplified Abelian EOM in the $U(1)$ thermal bath are:

\begin{equation}
\nabla\cdot\Evec = \rho_e,\qquad \nabla\times\Bvec = \Jvec
\end{equation}

\subsection*{Derivation of $\Jvec$}

To evaluate $\Jvec = \nabla \times B$, we start with Eq.~\ref{eq:W-curl}:

\begin{equation}
\nabla\times\Wvec = -2\uvec - (\rvec\cdot\nabla)\uvec
\end{equation}

Taking the curl:

\begin{equation}
\nabla\times(\nabla\times\Wvec) = -2\,\nabla\times\uvec - \nabla\times\bigl((\rvec\cdot\nabla)\uvec\bigr)
\end{equation}

Now $\nabla\times\uvec = \omvec$.  For the second term, evaluate componentwise.  

\begin{equation}
[\nabla\times((\rvec\cdot\nabla)\uvec)]_i = \epsilon_{ijk}\partial_j\bigl(r_\ell\partial_\ell u_k\bigr) = \epsilon_{ijk}(\partial_j r_\ell)(\partial_\ell u_k) + \epsilon_{ijk}r_\ell\partial_j\partial_\ell u_k
\end{equation}

Since $\partial_j r_\ell = \delta_{j\ell}$, the first term becomes $\epsilon_{ijk}\partial_j u_k = (\nabla\times\uvec)_i = \omega_i$.  The second term vanishes because $\partial_j\partial_\ell$ is symmetric while $\epsilon_{ijk}$ is antisymmetric in $j,k$ (after contraction with $r_\ell$).  Hence we have:

\begin{equation}
\nabla\times((\rvec\cdot\nabla)\uvec) = \omvec
\end{equation}

Therefore,

\begin{equation}
\nabla\times(\nabla\times\Wvec) = -2\omvec - \omvec = -3\omvec
\end{equation}

But this is only the linear part.  The full $\nabla\times\Bvec$ also contains the non‑Abelian contribution $\frac{g}{2}\nabla\times(\Wvec\times\Wvec)$.  However, a careful calculation (using the identity $\nabla\times(\avec\times\bvec) = \avec(\nabla\cdot\bvec) - \bvec(\nabla\cdot\avec) + (\bvec\cdot\nabla)\avec - (\avec\cdot\nabla)\bvec$) shows that the leading term in the current is:

\begin{equation}
\Jvec = -3\omvec - (\rvec\cdot\nabla)\omvec + \mathcal{O}(g)
\end{equation}

For the purposes of the dictionary in the main text, we keep only the linearised expression, which is sufficient to identify $\Jvec$ with vorticity and its gradient.  The full non‑Abelian corrections are higher order in $g$ and become important only inside the monopole cores.

\subsection*{Derivation of $\rho_e$}

Starting from $\Evec = -\nabla W_0 - \partial_0\Wvec$ (restoring $W_0$), Gauss’s law gives:

\begin{equation}
\rho_e = \nabla\cdot\Evec = -\nabla^2 W_0 - \partial_0(\nabla\cdot\Wvec)
\end{equation}

Using $W_0 = -H$ and $\nabla\cdot\Wvec = \nabla\cdot(\eta^{-2}\rvec\times\uvec) = -\eta^{-2}\rvec\cdot\omvec$ (since $\nabla\cdot(\rvec\times\uvec) = -\rvec\cdot\omvec$), we obtain:

\begin{equation}
\rho_e = \nabla^2 H + \eta^{-2}\,\partial_0(\rvec\cdot\omvec)
\end{equation}

The pressure Poisson equation for incompressible flow gives $\nabla^2 H = \nabla\cdot(\uvec\times\omvec)$.  Hence:

\begin{equation}
\rho_e = \nabla\cdot(\uvec\times\omvec) + \eta^{-2}\,\partial_0(\rvec\cdot\omvec)
\end{equation}

In the coarse‑grained description, $\eta^{-2}$ is absorbed into the definition of the fields, so the final expression in natural units (where $\eta=1$) is:

\begin{equation}
\rho_e = \nabla\cdot(\uvec\times\omvec) + \partial_t(\rvec\cdot\omvec)
\label{eq:rho-e1}
\end{equation}

This confirms the identification in the main text.  The first term is the divergence of the Lamb vector, and the second term is the time derivative of the radial component of vorticity.

\subsection*{The pressure Poisson equation from Gauss's law}
\label{sec:pressure-poisson}

A key consistency check between the gauge theory and classical fluid mechanics is the recovery of the pressure Poisson equation.  We start from the definition of the Bernoulli head:

\begin{equation}
H = \frac{p}{\rho} + \frac12 |\uvec|^2
\end{equation}

For an incompressible, inviscid flow, the Euler equation can be written in rotational form:

\begin{equation}
\partial_t \uvec + \nabla H = \uvec \times \omvec, \qquad \omvec = \nabla\times\uvec. \tag{Euler-rot}
\end{equation}

Taking the divergence of both sides gives:

\begin{equation}
\nabla^2 H = \nabla\cdot(\uvec \times \omvec) - \partial_t(\nabla\cdot\uvec)
\end{equation}

Since the flow is incompressible, \(\nabla\cdot\uvec = 0\), the time derivative term vanishes.  Hence we get:

\begin{equation}
\nabla^2 H = \nabla\cdot(\uvec \times \omvec) 
\label{eq:hydrodynamic-charge}
\end{equation}

This is the classical pressure Poisson equation, which determines the pressure from the velocity field.

In the gauge theory, Gauss's law for the massless \(U(1)\) sector is:

\begin{equation}
\nabla\cdot\Evec^3 = \rho_e^3
\end{equation}

with \(\Evec^3 = -\nabla W_0^3 - \partial_0\Wvec^3\) and \(W_0^3 = -H\).  From Appendix~E, the charge density is:

\begin{equation}
\rho_e^3 = \nabla\cdot(\uvec\times\omvec) + \partial_t(\rvec\cdot\omvec)
\end{equation}

Substituting these into Gauss's law gives:

\begin{equation}
-\nabla^2 H - \partial_0(\nabla\cdot\Wvec^3) = \nabla\cdot(\uvec\times\omvec) + \partial_t(\rvec\cdot\omvec)
\end{equation}

Using \(\Wvec^3 = \eta^{-2}\,\rvec\times\uvec\), we have \(\nabla\cdot\Wvec^3 = -\eta^{-2}\,\rvec\cdot\omvec\).  The time derivatives then cancel exactly, leaving:

\begin{equation}
\nabla^2 H = \nabla\cdot(\uvec\times\omvec)
\label{eq:bernoulli}
\end{equation}

which is precisely the pressure Poisson equation Eq.~\ref{eq:hydrodynamic-charge}.  Thus the gauge theory reproduces the classical pressure equation, providing a strong consistency check between the two descriptions.

\newpage

\section*{Appendix F: Derivation of \(\nabla\times\Wvec^3 = \omvec\)}

We derive the key identification used in the Wilson loop derivation and elsewhere: in the broken phase, the curl of the massless gauge field equals the fluid vorticity.\\

Start from the Helmholtz decomposition of the full gauge field in the broken phase:

\begin{equation}
\Wvec = -\nabla\Phi_L + \nabla\times\AL
\end{equation}

The massive sector \(W^{1,2}\) is confined to the vortex cores and decays exponentially outside.  In the bulk (far from any defect), only the massless \(U(1)\) component survives.  Absorbing the normalization factor \(g^2\) into the definition of \(\AL\), we write:

\begin{equation}
\Wvec^3 = \nabla\times\AL
\label{eq:W3-AL}
\end{equation}

The gradient part \(-\nabla\Phi_L\) does not contribute to the curl, so it is irrelevant for the vorticity identification.

For a filament aligned with the \(z\)-axis, the vector potential is azimuthal: \(\AL = A_\varphi(R)\,\hat{\varphi}\).  Then we have:

\begin{equation}
\nabla\times\AL = \frac{1}{R}\frac{\partial}{\partial R}(R A_\varphi)\,\hat{z} \equiv B_z(R)\,\hat{z}
\end{equation}

Thus,

\begin{equation}
\Wvec^3 = B_z(R)\,\hat{z}
\end{equation}

Taking the curl of \(\Wvec^3\):

\begin{equation}
\nabla\times\Wvec^3 = \nabla\times(B_z \hat{z}) = -\frac{1}{R}\frac{\partial B_z}{\partial R}\,\hat{z}
\label{eq:curl-W3-1}
\end{equation}

Now recall the reconstruction of the solenoidal velocity from the vector potential (using Eq.~\ref{eq:uA} and $\Lvec_A  = \nabla \times \AL$):

\begin{equation}
\uvec_A = \frac{1}{R^2}(\nabla\times\AL)\times\rvec
\label{eq:AL-to-u_A}
\end{equation}

where \(R\) is the cylindrical radius from the filament axis and \(\rvec = R\hat{R}\).  Using \(\nabla\times\AL = B_z(R)\,\hat{z}\), we have:

\begin{equation}
\uvec_A = \frac{1}{R^2}(B_z\hat{z})\times(R\hat{R}) = -\frac{B_z}{R}\,\hat{\varphi}
\end{equation}

This is the expected azimuthal swirling flow around the filament.  Taking the curl in cylindrical coordinates:

\begin{equation}
\nabla\times\uvec_A = \frac{1}{R}\frac{\partial}{\partial R}(R\,u_\varphi)\,\hat{z}
= \frac{1}{R}\frac{\partial}{\partial R}\left(R\left(-\frac{B_z}{R}\right)\right)\hat{z}
= -\frac{1}{R}\frac{\partial B_z}{\partial R}\,\hat{z}
\label{eq:curl-W3-2}
\end{equation}

Comparing Eq.~\ref{eq:curl-W3-1} and Eq.~\ref{eq:curl-W3-2}, we obtain:

\begin{equation}
\nabla\times\Wvec^3 = \nabla\times \uvec_A
\end{equation}

Now  $\nabla \times \uvec_A = \omvec_A$ as defined previously.  But considering Eq.~\ref{eq:totalvelocity}, within the thermal bath, $\omvec_\Phi \approx 0$ and also since $\uvec_r \approx v$, $\omvec_r \approx 0$. Therefore, we can write $\omvec_A = \omvec$ within the bath allowing us to write:

\begin{equation}
\nabla\times\Wvec^3 = \omvec
\end{equation}

\newpage

\newpage



\begin{thebibliography}{99}

\bibitem{VincentMeneguzzi1991} A. Vincent and M. Meneguzzi, ``The spatial structure and statistical properties of homogeneous turbulence,'' J. Fluid Mech. \textbf{225}, 1 (1991).

\bibitem{Jimenez1993} J. Jiménez, A. A. Wray, P. G. Saffman, and R. S. Rogallo, ``The structure of intense vorticity in isotropic turbulence,'' J. Fluid Mech. \textbf{255}, 65 (1993).

\bibitem{Arnold1966} V.~I. Arnold, ``Sur la géométrie différentielle des groupes de Lie de dimension infinie et ses applications à l'hydrodynamique des fluides parfaits,'' Ann. Inst. Fourier \textbf{16}, 319 (1966).

\bibitem{Ishihara2009} T. Ishihara, T. Gotoh, and Y. Kaneda, ``Study of high-Reynolds number isotropic turbulence by direct numerical simulation,'' Annu. Rev. Fluid Mech. \textbf{41}, 165 (2009).

\bibitem{FarooqBeyond2026}
Farooq, A. (2026)
Beyond Vorticity: An Angular Momentum Perspective on Fluid Flow.
\textit{arXiv:2605.21191}

\bibitem{FarooqIsotropic2026}
Farooq, A. (2026)
A Statistical Field Theory for Isotropic Turbulence.
\textit{arXiv:2604.19458}

\bibitem{lee2012spatiallydevelopingturbulentboundary}
J.~H.~Lee, Y.~S.~Kwon, N.~Hutchins, and J.~P.~Monty,
\newblock \emph{Spatially developing turbulent boundary layer on a flat plate},
\newblock arXiv:1210.3881 [physics.flu-dyn] (2012),
\newblock \url{https://arxiv.org/abs/1210.3881}.

\bibitem{marsden1983reduction}
Marsden, J. E., and Weinstein, A. (1983).
Coadjoint orbits, vortices, and Clebsch variables for incompressible fluids.
\textit{Physica D: Nonlinear Phenomena}, 7(1-3), 305-323.

\bibitem{polyakov1995two}
Polyakov, A. M. (1995).
Two-dimensional turbulence.
\textit{Nuclear Physics B}, 438(1-2), 255-276.

\bibitem{migdal1995loop}
Migdal, A. A. (1995).
Loop equation and area law in turbulence.
In \textit{Quantum Field Theory and String Theory} (pp. 193-218). 
Springer, Boston, MA.

\bibitem{moffatt1969degree}
Moffatt, H. K. (1969).
The degree of knottedness of tangled vortex lines.
\textit{Journal of Fluid Mechanics}, 35(1), 117-129.

\bibitem{moffatt1990knot}
Moffatt, H. K., and Ricca, R. L. (1990).
Helicity and the Calugareanu invariant.
\textit{Proceedings of the Royal Society of London. Series A: Mathematical and Physical Sciences}, 439(1906), 411-429.

\bibitem{freedman1991non}
Freedman, M. H., and He, Z. X. (1991).
Divergence-free fields: Energy and asymptotic crossing number.
\textit{Annals of Mathematics}, 134(1), 189-229.

\bibitem{jackiw2002noncommutative}
Jackiw, R., Nair, V. P., and Polychronakos, A. P. (2002).
Chern-Simons reduction and non-Abelian fluid mechanics.
\textit{Physical Review D}, 66(4), 045019.

\bibitem{holm1998euler}
Holm, D. D., Marsden, J. E., and Ratiu, T. S. (1998).
The Euler-Poincaré equations and semidirect products with applications to continuum theories.
\textit{Advances in Mathematics}, 137(1), 1-81.

\bibitem{ForsterNelsonStephen1977}
D.~Forster, D.~R.~Nelson, and M.~J.~Stephen, 
``Large-distance and long-time properties of a randomly stirred fluid,'' 
Phys. Rev. A \textbf{16}, 732 (1977).

\bibitem{YakhotOrszag1986}
V.~Yakhot and S.~A.~Orszag, 
``Renormalization-group analysis of turbulence,'' 
Phys. Rev. Lett. \textbf{57}, 1722 (1986).

\bibitem{Eyink1994}
G.~L.~Eyink, 
``The renormalization group method in statistical hydrodynamics,'' 
Phys. Fluids \textbf{6}, 3063 (1994).

\bibitem{McCombWatt1992}
W.~D.~McComb and A.~G.~Watt, 
``Two-field theory of incompressible-fluid turbulence,'' 
Phys. Rev. A \textbf{46}, 4797 (1992).

\bibitem{Verma2025}
M.~Verma, 
``Turbulence: A Nonequilibrium Field Theory,'' 
arXiv:2501.19367 (2025).

\bibitem{khesin2009topology}
Khesin, B., and Wendt, R. (2009).
\textit{The geometry of infinite-dimensional groups}.
Springer Science \& Business Media.

\bibitem{Peskin-1995}
M.~E.~Peskin and D.~V.~Schroeder,
\newblock {\em An Introduction to Quantum Field Theory},
\newblock Addison-Wesley, Reading, USA, 1995.

\bibitem{HolmKupershmidt1983}
D. D. Holm and B. A. Kupershmidt,
“Yang–Mills fluids and their symplectic structure,”
\emph{Phys. Lett. A} \textbf{97}, 133–136 (1983).

\bibitem{Barducci1985}
A. Barducci, R. Casalbuoni, and L. Lusanna,
“Odd variables in classical mechanics and in Yang–Mills fluid dynamics,”
\emph{Phys. Lett. A} \textbf{112}, 271–272 (1985).

\bibitem{Jackiw2002}
R. Jackiw, S.-Y. Pi, and A. P. Polychronakos,
“Noncommuting gauge fields as a Lagrange fluid,”
\emph{Ann. Phys.} \textbf{301}, 157–173 (2002).

\bibitem{Kambe2003}
T. Kambe,
“Gauge principle for flows of an ideal fluid,”
\emph{Fluid Dyn. Res.} \textbf{32}, 193–199 (2003).

\bibitem{Kambe2007}
T. Kambe,
“Gauge principle and variational formulation for ideal fluids with reference to translation symmetry,”
\emph{Fluid Dyn. Res.} \textbf{39}, 98–117 (2007).

\bibitem{Marmanis1998}
H. Marmanis,
“Analogy between the electromagnetic and hydrodynamic equations: Application to turbulence,”
\emph{Phys. Fluids} \textbf{10}, 1428–1437 (1998).

\bibitem{Mendes2006}
A. C. R. Mendes, C. Neves, W. Oliveira, and F. I. Takakura,
“Noncommutative metafluid dynamics,”
\emph{Int. J. Mod. Phys. A} \textbf{21}, 505–516 (2006).

\bibitem{Peshkov2019}
I. Peshkov, E. I. Romenski, and M. Dumbser,
“Continuum mechanics with torsion,”
\emph{Continuum Mechanics and Thermodynamics} \textbf{31}, 1741–1765 (2019)

\bibitem{tHooft1974} G. 't Hooft, ``Magnetic monopoles in unified gauge theories,'' Nucl. Phys. B \textbf{79}, 276 (1974).

\bibitem{Polyakov1974} A. M. Polyakov, ``Particle spectrum in quantum field theory,'' JETP Lett. \textbf{20}, 194 (1974).

\bibitem{Bogomolnyi1976} E. B. Bogomol'nyi, ``Stability of classical solutions,'' Sov. J. Nucl. Phys. \textbf{24}, 449 (1976) [Yad. Fiz. \textbf{24}, 861 (1976)].

\bibitem{PrasadSommerfield1975} M. K. Prasad and C. M. Sommerfield, ``Exact classical solution for the 't Hooft monopole and the Julia-Zee dyon,'' Phys. Rev. Lett. \textbf{35}, 760 (1975).

\bibitem{WuLambDivergence}
J.~Z.~Wu, ``Lamb vector divergence and hydrodynamic charge density,'' arXiv:1206.1281 (2012).

\bibitem{ZLF1992} V.~E. Zakharov, V.~S. L'vov, and G.~Falkovich, \textit{Kolmogorov Spectra of Turbulence I: Wave Turbulence}, Springer Series in Nonlinear Dynamics (Springer-Verlag, Berlin, Heidelberg, 1992).

\bibitem{Li2008} Y. Li, E. Perlman, M. Wan, Y. Yang, C. Meneveau, R. Burns, S. Chen, A. S. Szalay, and G. L. Eyink, ``A public turbulence database cluster and applications to study Lagrangian evolution of velocity increments,'' J. Turbul. \textbf{9}, N31 (2008).

\bibitem{Migdal1994} A. A. Migdal, ``Loop equation and area law in turbulence,'' Int. J. Mod. Phys. A \textbf{9}, 1197 (1994).

\bibitem{Pirozzoli2012} S. Pirozzoli, ``On the velocity and dissipation signature of vortex tubes in isotropic turbulence,'' Physica D: Nonlinear Phenomena Volume 241, Issue 3, (2012).

\bibitem{VainshteinSreenivasan2004} S. I. Vainshtein and K. R. Sreenivasan, ``On the distribution of dissipation in turbulent flows,'' Phys. Fluids \textbf{16}, 4259 (2004).

\bibitem{MoisyJimenez2004} F. Moisy and J. Jiménez, ``Geometry and clustering of intense structures in isotropic turbulence,'' J. Fluid Mech. \textbf{513}, 111 (2004).

\bibitem{Frisch1995}
U.~Frisch, \textit{Turbulence: The Legacy of A. N. Kolmogorov} (Cambridge University Press, 1995).

\end{thebibliography}
\end{document}